\documentclass{aa}

\usepackage{natbib}
\bibpunct{(}{)}{;}{a}{}{,} 

\usepackage{txfonts}

\usepackage{supertabular}
\usepackage{threeparttable}
\usepackage{array,longtable}
\usepackage{multirow}
\usepackage{multicol}
\usepackage{booktabs}
\usepackage{amsmath}
\usepackage{gensymb}
\usepackage{graphicx}
\graphicspath{{figs/}}
 \usepackage{float}
\usepackage{verbatim}
\usepackage{amssymb}
\usepackage{lineno}
\usepackage{wasysym}
\usepackage{amssymb}
\usepackage{xcolor}
\usepackage{soul}
\usepackage{amstext}
\usepackage{array}
\usepackage{subfig}
\usepackage{enumitem}
\usepackage{amssymb}
\usepackage{caption}
\setlist[enumerate,1]{label={\arabic*.}}
\setlist[enumerate,2]{label={(\alph*)}}
\newcommand\repeatcaption{%
  \if@firstcolumn\multicolumn{4}{c}{Table \thetable: some caption (Continued)}\\[10pt]\fi}
\makeatother

\usepackage[T1]{fontenc} 
\usepackage{lmodern} 
\rmfamily 
\DeclareFontShape{T1}{lmr}{b}{sc}{<->ssub*cmr/bx/sc}{}
\DeclareFontShape{T1}{lmr}{bx}{sc}{<->ssub*cmr/bx/sc}{}

\newcommand{\kms}{km~s$^{-1}$}

\newcommand{\mgii}{\ion{Mg}{ii}}
\newcommand{\civ}{\ion{C}{iv}}

\newcommand{\icm}{cm$^{-2}$}

\newcommand{\za}{$z_{\rm abs}$}

\newcommand{\dcthree}{\mbox{$\langle d_c^{3D} \rangle$}}
\newcommand{\dctwo}{\mbox{$\langle d_c^{2D} \rangle$}}

\newcommand{\str}{\mbox{$\sigma_\perp$}}
\newcommand{\spa}{\mbox{$\sigma_\parallel$}}
\newcommand{\mspa}{\mbox{$\langle \sigma_\parallel\rangle$}}

\newcommand{\xiq}{\mbox{$\xi^{QSO}(\Delta v)$}}
\newcommand{\xia}{\mbox{$\xi^{arc}(\Delta v)$}}

\newcommand{\tpcf}{$\xi(\Delta r_\perp,\Delta v)$}

\newcommand{\one}{1527 a}
\newcommand{\two}{1527 b}
\newcommand{\three}{1527 c}
\newcommand{\four}{G311 a}
\newcommand{\five}{0033 a}
\newcommand{\six}{0033 b}
\newcommand{\seven}{2111 a}
\newcommand{\eight}{2111 b}

\begin{document}

\title{Transverse clues on the kiloparsec-scale structure of the
  circumgalactic medium as traced by C\,IV absorption}

\author{
  S. Lopez,\inst{1}
  A. Afruni,\inst{1,2,3}
 D. Zamora,\inst{1}
{N. Tejos,}\inst{4}
{C. Ledoux,}\inst{5}
{J. Hernandez,}\inst{6} 
{T.~A.~M. Berg,}\inst{7}
{H. Cortes,}\inst{1}
{F. Urbina, }\inst{1}
{E. J. Johnston, }\inst{8}
{L. F. Barrientos,}\inst{6}
{M. B. Bayliss,}\inst{9}
{R. Cuellar,}\inst{1}
{J.~K. Krogager,}\inst{10,11}
{P. Noterdaeme,}\inst{10,12}
\and 
{M. Solimano}\inst{8}
}

\institute{Departamento de Astronom\'ia, Universidad de Chile,
  Casilla 36-D, Santiago, Chile. 
  \email{slopez@das.uchile.cl}
\and
Kapteyn Astronomical Institute, University of
Groningen, Landleven 12, 9747 AD Groningen, The Netherlands
\and
Dipartimento di Fisica e Astronomia, Universit\`a di Firenze, Via
G. Sansone 1, 50019 Sesto Fiorentino, Firenze, Italy 
\and
Instituto de F\'isica, Pontificia Universidad Cat\'olica de
Valpara\'iso, Casilla 4059, Valpara\'iso, Chile
\and
European Southern Observatory, Alonso de C\'ordova 3107, Vitacura, Casilla 19001, Santiago, Chile
\and 
Instituto de Astrofı\'isica, Pontificia Universidad Cat\'olica de
Chile, Av. Vicu\~na Mackenna 4860, 7820436 Macul, Santiago, Chile
\and 
NRC Herzberg Astronomy and Astrophysics Research Centre, 5071 West Saanich Road, Victoria, B.C., Canada, V9E 2E7
\and 
Instituto de Estudios Astrof\'isicos, Facultad de Ingenier\'ia y
Ciencias, Universidad Diego Portales, Av. Ej\'ercito Libertador 441,
Santiago, Chile
\and 
Department of Physics, University of Cincinnati, Cincinnati, OH 45221,
USA 
\and
French-Chilean Laboratory for Astronomy, IRL 3386, CNRS and U. de
Chile, Casilla 36-D, Santiago, Chile
\and
Centre de Recherche Astrophysique de Lyon, Université de Lyon 1,
ENS-Lyon, UMR5574, 9 Av Charles André, 69230 Saint-Genis-Laval, France
\and 
 Institute d’Astrophysique de Paris, CNRS-SU, UMR 7095, 98bis bd Arago, 75014 Paris, France
}

\titlerunning{\civ\ transverse kinematics}
\authorrunning{S. Lopez et al.}

\abstract{The kiloparsec-scale kinematics and density structure of the
  circumgalactic medium (CGM) is still poorly constrained
  observationally, which poses a problem for understanding the role of
  the baryon cycle in galaxy evolution. Here we present VLT/MUSE 
  integral-field spectroscopy ($R\approx 1\,800$) {of four giant
    gravitational arcs exhibiting $W_0 \gtrsim 0.2$
    \AA\ \civ\ absorption at eight intervening redshifts, $z_{abs}\approx
    2.0$--$2.5$.  We detected \civ\ absorption in a total of 222 adjacent and
    seeing-uncorrelated sight lines whose spectra sample beams of
    (``de-lensed'') linear size $\approx 1$ kpc.  Our data show that
    (1) absorption velocities cluster at all probed transverse scales,
    $\Delta 
    r_\perp\approx0$--$15$ kpc, depending on system; (2) the
    (transverse) velocity dispersion never exceeds the mean
    (line-of-sight) absorption spread; and (3) 
the (transverse) velocity autocorrelation function does not resolve
kinematic patterns at the above spatial scales, but its velocity
projection, \xia, exhibits a similar shape to the known two-point correlation
function toward quasars, \xiq. 
  } An empirical
  kinematic model suggests that these results are a natural
  consequence of wide-beam observations of an unresolved clumpy
  medium.  
  Our model recovers both the underlying velocity
  dispersion of the clumps ($70$--$170$ \kms) and the mean number of
  clumps per unit area ($2$--$13$ kpc$^{-2}$). The latter constrains
  the projected mean inter-clump distance to within
  $\approx0.3$--$0.8$ kpc, which we argue is a measure of clump size
  for 
  a  near-unity covering fraction.
  The model is also able to
  predict \xia\ from \xiq, suggesting that the strong systems that
  shape {  \xia\ and the line-of-sight velocity components that
    define \xiq\ } trace the same kinematic
  population. Consequently, the clumps must possess an internal
  density structure that {generates both weak and strong
    components.} We discuss how our interpretation is consistent with
  previous observations using background galaxies and multiple
  quasars as well as its implications for the connection between the
  small-scale kinematic structure of the CGM and galactic-scale 
  accretion and feedback processes.}


  \maketitle

\section{Introduction} 
\label{introduction}

The widespread presence of metals in the diffuse intergalactic and
circumgalactic media ~\citep[IGM and CGM, respectively;
  e.g.,][]{Cowie1995,Ellison2000,Simcoe2004,Songaila2005,Ryan-Weber2006,Becker2019,Cooper2019,DOdorico2023,Bordoloi2024}
suggests a continuous metal enrichment since $z > 4$ that remained
constant until $z = 2$~\citep{McQuinn2016,DOdorico2022,Galbiati2023}.
Produced in
galaxies~\citep[e.g.,][]{Pettini2001,Adelberger2005,Lofthouse2023,Banerjee2023}
and their environments~\citep{Shen2012}, CGM metals enter a baryon
cycle powered by galactic-scale feedback and re-accretion mechanisms
that are believed to regulate star formation and ultimately galaxy
evolution~\citep[e.g.,][]{Keres2005,Oppenheimer2008,Faucher-Ciguere2011,Christensen2016}.
Both the outflowing~\citep{Rupke2005,Oppenheimer2006} and  
inflowing~\citep[][]{Nelson2015,Fielding2017,Faucher-Ciguere2023} gas
may be clumpy, resulting in poor  mixing of the metals on ``small scales''~\citep[][]{Schaye2007}, defined here as $\lesssim 1$ kpc. Thus, {the CGM small-scale kinematics and
  spatial structure must be intimately connected to the baryon cycle
  of galaxies}. Constraining the former observationally has become a cornerstone for understanding the latter and validating   
hydro simulations of increasingly higher resolution~\citep[e.g.,][]{Oppenheimer2008,Wiersma2010,Cen2011,Rahmati2016,Bird2016,Finlator2020,Hummels2019,Peeples2019,Marra2024}.

Measuring the clumpiness of the CGM is difficult because 
this medium is diffuse, and the
rare bright background sources (e.g., quasars) required to detect it in
absorption do not provide transverse sampling. 
The only option to directly probe the transverse dimension
seems to rely on the even scarcer multiple background sources. In
fact, resolved spectroscopy of lensed quasars and galaxies already have
provided evidence that the $T=10^4$ K enriched gas is clumpy on kiloparsec
scales~\citep{Rauch1999,Rauch2001,Rauch2001a,Ellison2004,Lopez2005,
  Lopez2007,Chen2014,Zahedy2016,Rubin2018b,Rubin2018, Peroux2018a,
  Lopez2018,Krogager2018,Kulkarni2019,Zahedy2019,Lopez2020,Mortensen2021,Bordoloi2022,Afruni2023}.

An alternative way to probe the CGM structure on various scales is
through measuring line-of-sight and transverse velocity clustering. This
technique has been applied on different data sets, for instance, 
({ i}) the auto-correlation
function of absorbers either along the line-of-sight at high spectral
resolution~\citep{Steidel1990a,Pichon2003,Scannapieco2006,Boksenberg2015,Rauch1996,Fathivavsari2013}
or transversely using lensed or multiple quasars in
general~\citep{Rauch2001,Coppolani2006,Tytler2009,Martin2010,Mintz2022,Maitra2019,Gontcho2018,Dutta2024,Hennawi2006};
({ ii}) the absorber-galaxy cross-correlation either along the line-of-sight
using background galaxies~\citep{Steidel2010,Turner2017} or
transversely
~\citep{Adelberger2005,Lofthouse2023,Banerjee2023,Galbiati2023}; 
and ({ iii}) other cross-correlations such as quasar-absorber~\citep{Hennawi2007,Vikas2013} or outflow-absorber~\citep{Rauch2001a,Theuns2002}. The
broad picture that has emerged is that metals and galaxies trace the same
overdensities. However, none of the observations can really
disentangle the absorbing galaxy from the absorption system. Only a
handful of them address the line-of-sight
kinematics~\citep[e.g.][]{Turner2017} and how this is supposed to be entangled with the spatial
structure~\citep{Stern2016}, and even fewer have really been able
to { measure} the level of transverse 
structure~\citep{Rauch1999,Rauch2001, Rauch2001a}.

In this article, we take advantage of  multiplexed spectroscopy of
giant gravitational 
arcs (hereafter "ARCTOMO\footnote{\tt
  https://sites.google.com/view/arctomo/home} data")
to measure the velocity clustering of 
intervening triply ionized carbon (\civ) across kiloparsec scales. 
The strong and easy-to-identify \civ$\lambda\lambda 1548,1550$ doublet 
is arguably the most sensitive 
metal tracer of both the cool and warm CGM at high redshifts~\citep{Chen2001,Bordoloi2014,Rudie2019}. 
On the other hand, ARCTOMO data have demonstrated the potential to add a
wealth of unique and new spatial  
information on CGM 
scales~\citep{Lopez2018,Lopez2020,Tejos2021,Fernandez-Figueroa2022,
  Afruni2023}, hence the timely combination. We build on a {blind} survey of $z\approx 2$ \civ\  in all
available ARCTOMO fields (originally targeted for $z\approx 1$ \mgii) that  resulted  in eight intervening \civ\ systems toward four arcs.  

The paper is organized as follows. In Sect.~\ref{sect_data}, we describe
the observations, the data reduction, and the spectrum extraction. Then
in Sect.~\ref{sect_observables} we describe the automated
\civ\ identification and line-profile fitting used, paying special
attention to survey completeness, and in Sect.~\ref{sect_lensing} we address
the subtleties of dealing with 
lensed fields. We used these data for two kinds of analysis on
\civ\ kinematics: First, in Sect.~\ref{sect_direct}, we carry out a
direct assessment of line-of-sight and transverse kinematic
properties, and secondly, in Sect.~\ref{sect_correlation}, we measure
transverse velocity clustering and compare it with quasar
line-of-sight observations. Thereupon, in Sect.~\ref{sect_simple_model},
we present a gas-kinematics model that explains the results,
reproduces the arc signal out of the quasar kinematics, and predicts
independent observations. Finally, in~Sect.~\ref{sect_discussion}, we
discuss the implications of our findings, concluding with a summary of
the results in Sect.~\ref{sect_summary}.

Throughout the paper we use a $\Lambda$CDM cosmology with the
following cosmological parameters: $H_0=70$
\,km\,s$^{-1}$\,Mpc$^{-1}$, $\Omega_m=0.3$, and $\Omega_{\Lambda}
=0.7$. We also use the standard notations ${\cal N}(\mu,\sigma^2)$ for a normal
distribution with mean $\mu$ and variance $\sigma^2$ and ${\cal
  U}_{[a,b]}$ for a continuous uniform distribution with support
[a,b].

\section{ARCTOMO data}
\label{sect_data}

\begin{table*}
  \centering
  \small
\caption{Summary of targets and VLT/MUSE observations.}
\label{tab:obs}
\begin{tabular}{lcccccccc}
  \hline
\hline
Field name &  R. A. &  Dec.  &$z_{em}$ & Total Exp. &  Instrument & Effective  & Program ID & Lens\\
           &  [h:m:s] &  [\degree:\arcmin:\arcsec]  &&  time [s]      &    Mode    & PSF$^a$ [\arcsec] & & Ref.$^b$\\
\hline
SGAS\,J0033+0242 & 00:33:41.52 & +02:42:27.0 &2.39& $11 \times 700$ &  WFM-NOAO-N & 0.84 & 098.A-0459(A) & (1)\\
SGAS\,J1527+0652 & 15:27:44.74 & +06:52:19.4 &2.76&  $25 \times 600$  & WFM-AO-E & 0.78 & 0103.A-0485(B) & (2)\\
PSZ1\,G311.65-18.48 & 15:50:07.46 & -78:11:23.2 &2.37&  $2 \times 1483$ &   WFM-NOAO-N & 0.82 &  297.A-5012(A) & (3)\\
SGAS\,J2111-0114 & 21:11:19.01 & -01:14:30.4 &2.86&  $15 \times 606$  & WFM-AO-E & 0.74 & 0103.A-0485(B) & (4)\\
\hline

\end{tabular}
\begin{minipage}{0.99\textwidth}
\footnotesize{
$^a$ Point spread function at $\approx 5\,000$\,\AA\ measured from fitting 2D Gaussian profiles to stars in the final combined MUSE cubes. \\
$^b$ Lens model references: (1)~\citet{Fischer2019}, (2)~\citet{Sharon2020}, (3)~\citet{Lopez2020}, and (4)~\citet{Sharon2020}.

}
\vspace{2ex}
\end{minipage}
\end{table*}

\subsection{Observations}

The four arcs were observed  with the
Multi Unit Spectroscopic Explorer~\citep[MUSE; ][]{MUSE} mounted on
UT4 (Yepun) at the Very Large Telescope (VLT) in Paranal, Chile, as
part of programs 297.A-5012 (PI Aghanim), 098.A-0459 (PI Lopez), and
0103.A-048 (PI Lopez). The observations were conducted {in
service mode} with the
MUSE wide field mode (WFM), which provides a field of view of
$1\arcmin \times 1\arcmin$ {sampled at}  
0\farcs2/spaxel. They used either
non-adaptive optics and a nominal wavelength range (NOAO-N) or
adaptive optics and an extended wavelength range (AO-E), depending
on the program  (Table~\ref{tab:obs}). These setups provide spectral
coverage of $\approx 
4\,700$--$9\,300$\,\AA\ and $\approx 4\,600$--$9\,300$ 
\AA\, respectively,\footnote{A gap between 
  $5\,760$--$6\,010$\,\AA\ is present due to the contamination produced
  by the sodium laser used in AO.} and a resolving power ranging  
from \textit{R}$\simeq$1\,770 at 4\,800\,\AA\ to \textit{R}$\simeq$3\,590 at
9\,300\,\AA. Exposure times varied between $0.8$ and $4.2$ hours. 
  {The requested image quality 
  resulted in final 
  point spread functions (PSFs) ranging from $0\farcs74$ to $0\farcs84$}, 
depending on the targeted field.

\subsection{Data reduction}

The data reduction was carried out using the ESO MUSE pipeline
\citep[v2.6,][]{Weilbacher2020} in the ESO Recipe Execution Tool
(EsoRex) environment \citep{ESOREX2015}. The master bias, flat field,
and wavelength calibrations for each CCD were created from the
associated raw calibrations, and they were applied to the raw science and
standard-star observations as part of the pre-processing steps. Flux
calibration was carried out using the standard star observations from
the same nights as the science data, and the sky continuum was
measured directly from the science exposures and subtracted off. The
reduced data for each exposure were then stacked to produce the
combined 
datacube, with a wavelength solution calibrated to vacuum. 
Any
residual sky contamination was removed using the Zurich Atmosphere
Purge code \citep[ZAP, ][]{Soto2016}. Finally, the datacubes were matched to
the WCS of the corresponding HST data (references in
Table~\ref{tab:obs}).  White-image stamps of the arcs are shown in
the left-hand column of Fig.~\ref{fig_lensed}.

\subsection{Binned spectra extraction}

  We extracted and combined the cube spectra optimally using
  3$\times$3 spaxel apertures (so  binned spaxels are
  non-overlapping squares of 0\farcs6 on a side), a size deemed
  sufficiently large to 
  minimize cross-talk between adjacent spaxels due to seeing smearing,
  while maximizing spatial sampling~\citep[e.g., ][]{Lopez2018}. 
  {We show in Appendix~\ref{sect_sampling} that this choice does
    not bias our results.
    
  We considered 
  an arc binned spectrum to sample an independent ``arc
  sight line'' or ``arc beam.''}  By construction, these spectra have
  heterogeneous signal-to-noise ratios.  {Throughout the
    article, we apply appropriate completeness corrections to deal with
    each arc's particular selection function.}

\section{Absorption line analysis and definitions}
\label{sect_observables}

\begin{figure*}
  \centering 
  \includegraphics[width=1.7\columnwidth]{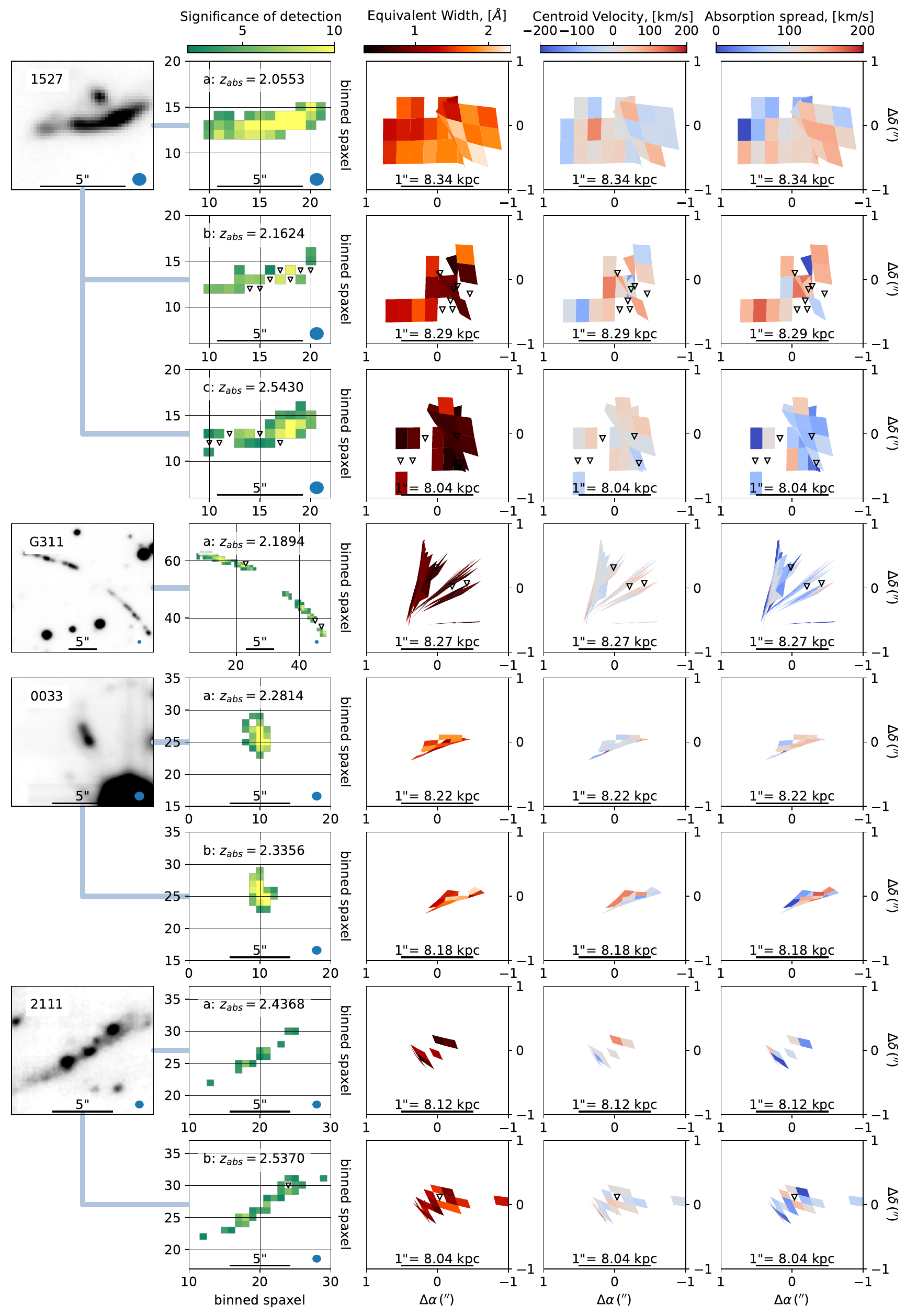}  
  \caption{
    MUSE arc images and \civ\ system maps. 
    Columns from left to right: 
Column (1):  White-image stamps of the gravitational arcs.   
Column (2):  $0\farcs6\times0\farcs6$-binned spaxels with  
    \civ\ detections, colored according to their significance (Sect.~\ref{sect_fitting}).  
    Spaxel coordinates as in
    Figs. A.1 to A.8.
    In both columns, the PSF is represented by the lower-right circle. 
Columns (3), (4), and (5): Absorber-plane 
    reconstructed maps of equivalent width ($W_0^{1548}$),  
    velocity, and absorption spread, respectively. 
    In all panels, the scale is indicated.  
    In Columns (2)--(5), the triangles indicate $W_0\le 0.2$
    \AA\ (2-$\sigma$) non-detections. 
  } \label{fig_lensed}
\end{figure*}

\begin{figure}
  \centering
  \includegraphics[
    width=1\columnwidth,clip]{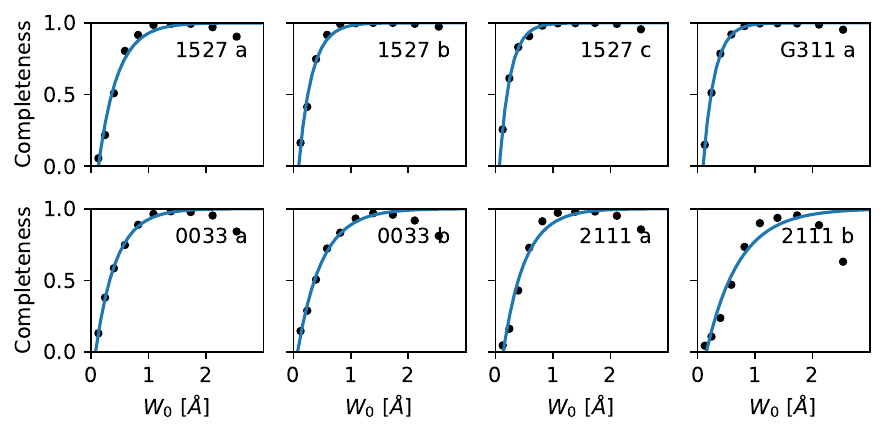}
    \includegraphics[width=1\columnwidth]{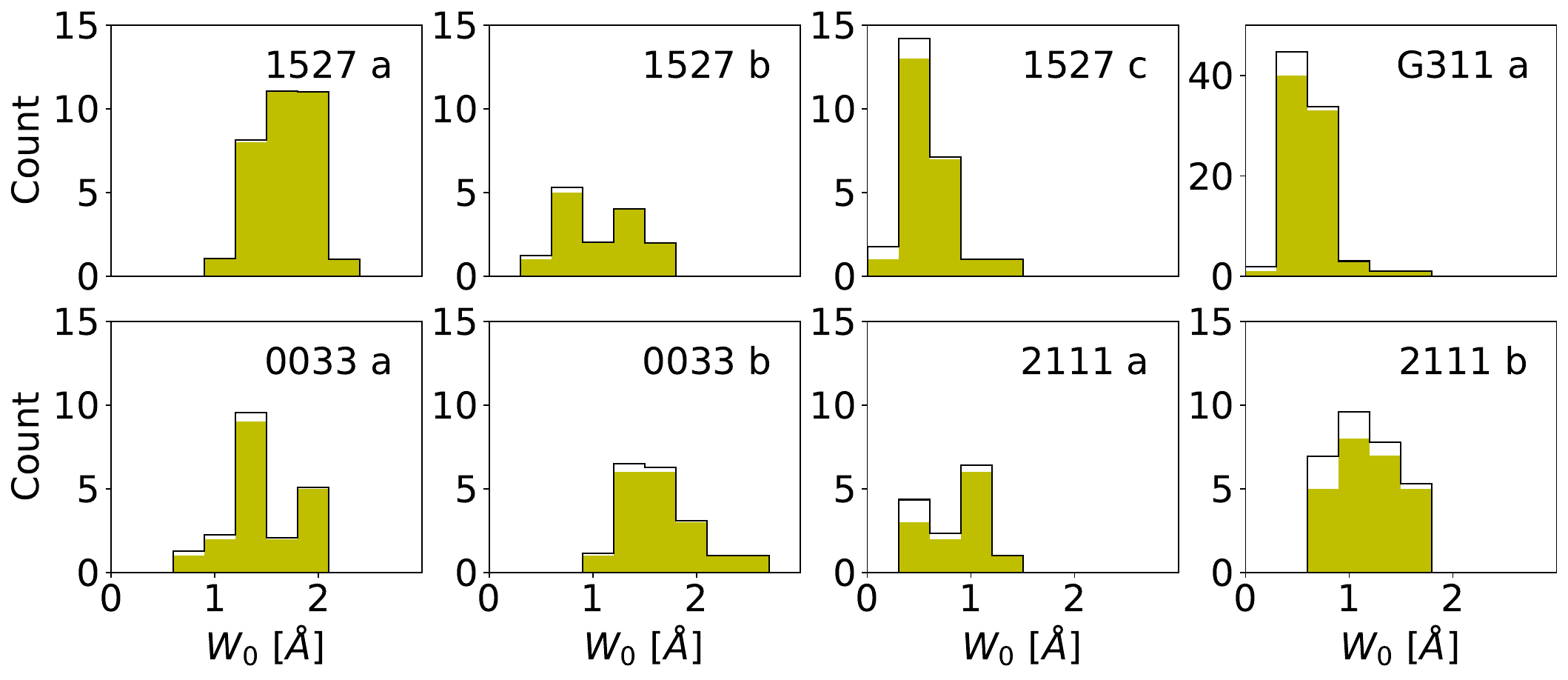} 
    \caption{
Rest-frame equivalent width distributions.  
      Upper panels: Completeness of $W_0$ and exponential
    fit. $W_0 \gtrsim 2$ 
    \AA\ produces blended doublet lines and are not used in the
    fit. Lower panels: Measured and
    completeness-corrected $W_0$ distributions 
    (filled and 
    unfilled histograms, respectively). We note that system \four\ has a different $y$-scale. }
  \label{fig_FNR}
\end{figure}

We defined a \civ\ ``ARCTOMO system,'' or simply a ``\civ\ system,'' as
having significant \civ\ absorption detected in at least one arc
sight line. We borrowed the concept of ``system'' from
the quasar 
absorption lines technique despite the fact that here we deal with
extended sources {so that a system could be probed by several adjacent
sight lines.} In this article, we refer to intervening systems,
$c|z_{abs} - z_{em}|/(1+z_{em})>3\,000$ \kms, {where $z_{abs}$
  and $z_{em}$ are the system absorption and  
  the source emission redshifts, respectively.
The \civ\ system identification and line profile fitting were performed separately and automatically. We outline these two steps below.}

\subsection{\civ\ system identification}
\label{sect_survey} 

To automatically find candidate systems, binned spectra were
pre-selected within a manually selected mask that contained the arc and
sufficient sky around it. We identified candidate \civ\ doublets using 
{the template-matching algorithm described
  in~\citet[][]{Noterdaeme2010} and~\citet{Ledoux2015}. The search
  runs over all masked spectra along the available redshift path (see
Appendix~\ref{sect_effectiveness} for details).} A total of eight
candidate \civ\ systems were found toward four arcs at redshifts \za\ between
$2$ and $2.5$.

\subsection{Line profile fitting}
\label{sect_fitting}

\begin{table*}
\centering
\begin{threeparttable}
\caption{Summary of \civ\ systems.}
\begin{tabular}{lcccccccccc}
\hline
\hline
system&$z_{abs}$&\# of&\# of&$\langle {\mathrm S/N} \rangle$&$\sigma_\perp$&$\langle\sigma_\parallel\rangle$&$W_0$-range&$\Delta r_\perp$-range&Area\\
&&spec.&det.&&[km/s]&[km/s]&[\AA]&[kpc]&[kpc$^2$]\\
&(1)&(2)&(3)&(4)&(5)&(6)&(7)&(8)&(9)\\\hline
1527 a&2.0553&45&32&6.9&40.6$\pm$4.0&131.5$\pm$2.9&1.2--2.2&0.2--13.9&117.6\\
1527 b&2.1624&49&14&8.8&65.7$\pm$5.7&132.8$\pm$6.8&0.4--1.7&0.7--12.3&52.1\\
1527 c&2.5430&55&23&10.2&22.4$\pm$3.4&61.7$\pm$4.4&0.2--1.4&0.2--14.4&65.5\\
G311 a&2.1894&181&79&9.5&30.4$\pm$2.1&62.0$\pm$3.1&0.3--1.5&0.0--10.5&39.5\\
0033 a&2.2814&50&19&7.4&74.6$\pm$5.2&134.7$\pm$5.5&0.7--1.9&0.1--8.7&9.1\\
0033 b&2.3356&50&18&7.6&57.7$\pm$4.9&127.4$\pm$5.6&1.0--2.6&0.1--6.4&9.3\\
2111 a&2.4367&50&12&5.1&64.9$\pm$7.2&86.4$\pm$8.4&0.4--1.3&0.2--4.4&5.9\\
2111 b&2.5370&53&25&3.6&54.4$\pm$5.7&83.3$\pm$6.3&0.7--1.7&0.1--9.6&15.4\\
\hline
Total&&533&222&&&&&&314.4\\
 \hline
\label{table_summary}
\end{tabular}
\begin{tablenotes}[flushleft]
\item Table columns: 
(1) Absorption redshift,
(2) {number} of binned spectra surveyed,
(3) {number} of \civ\ detections,
(4) median pixel S/N,
(5) transverse velocity dispersion,
(6) median absorption spread,
(7) range of \civ\ rest-frame equivalent widths,
(8) range of transverse distances between spaxel centers, and
(9) total area {of \civ\ detections.} 
\end{tablenotes}
\end{threeparttable}
\end{table*}

\begin{table}
\centering
\begin{threeparttable}
\caption{Ancillary data.}
\vspace{1mm}
\centering
\begin{tabular}{cccccc}
        \toprule
        \multirow{2}{*}{System} & 
        \multicolumn{1}{c}{$\lambda^{1548}$} & 
        \multicolumn{1}{c}{$\langle \delta_v \rangle$} & 
        \multicolumn{1}{c}{$\sigma_{\textrm{inst}}$} & 
        \multicolumn{1}{c}{$\sigma_\perp^{\rm MC}$} &
        \multirow{2}{*}{$\alpha$} \\ 
        & \multicolumn{1}{c}{[\AA]} &
        \multicolumn{1}{c}{[km/s]} & 
        \multicolumn{1}{c}{[km/s]} & 
        \multicolumn{1}{c}{[km/s]} & \\
         & (1) & (2) & (3) & (4) & (5) \\
        \midrule
    1527 a & 4730.2 & 17.5 & 67.2 & 17.2 & -3.0 \\
    1527 b & 4896.0 & 20.8 & 64.9 & 27.2 & -3.9 \\
    1527 c & 5485.3 & 12.2 & 57.9 & 18.3 & -5.4 \\
    G311 a & 4937.8 & 14.4 & 64.3 & 19.8 & -5.1 \\
    0033 a & 5080.2 & 21.0 & 62.5 & 24.9 & -2.3 \\
    0033 b & 5164.2 & 17.8 & 61.5 & 19.2 & -2.0 \\
    2111 a & 5320.7 & 22.8 & 59.7 & 28.8 & -2.6 \\
    2111 b & 5476.0 & 20.2 & 58.0 & 22.1 & -1.6 \\
    \bottomrule
    \label{table_corrections}
\end{tabular}
\begin{tablenotes}[flushleft]
   \footnotesize
\item Table columns: 
(1) Wavelength of \civ\ zero-point velocity;
(2) median velocity uncertainty (the
individual velocity errors $\delta_v$ are used in Sect.~\ref{sect_direct}
to carry out the bootstrap analysis on \str\, and in
Sect.~\ref{sect_correlation} to produce $v=0$ mock catalogs.
 These are
listed in Table~A.1 and their distributions displayed
in Fig.~\ref{fig_errors});
(3) instrumental profile width at $\lambda^{1548}$ ($\sigma_{\rm inst}$ is used to
compute \spa\ in Sect.~\ref{sect_direct}. $\sigma_{\rm inst}$ was
obtained at each system's wavelength by interpolating the values
measured by~\citet{Mentz2016} using sky lines);
(4) spurious \str\ (Sect.~\ref{sect_direct} and Appendix~\ref{sect_spurious_dispersion});
(5) slope of exponential fit to $W_0$-completeness
(Sect.~\ref{sect_completeness}). 
\end{tablenotes}
\end{threeparttable}
\end{table}

\begin{figure*}
  \centering
  \includegraphics[width=2\columnwidth]{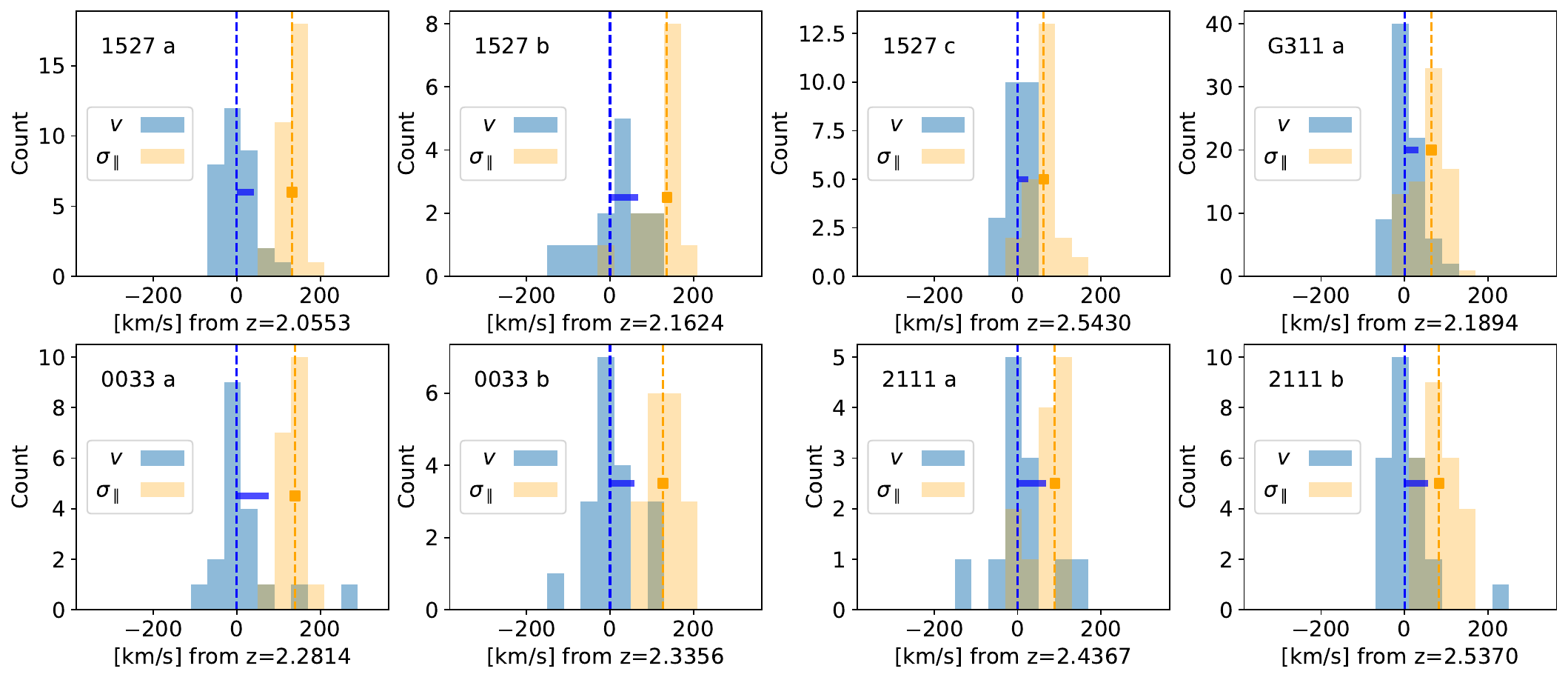}  
  \caption{Distribution of \civ\ doublet velocity ($v$) and absorption
    spread (\spa) by system. The blue dashed line marks the
    zero-point velocity, and the orange
    dashed line indicates the median absorption spread ($\langle
    \spa\rangle$).  The blue horizontal solid line has a size equal to
    the transverse velocity dispersion (\str); its blue side is
    arbitrarily placed at zero velocity for a visual comparison between \str\ and $\langle
    \spa\rangle$. 
    The data
    represented in this figure has not been corrected for
    incompleteness.  } \label{fig_velocities}
\end{figure*}

In a second round, an automated {fit} of a double Gaussian plus a
local continuum is ran over all masked spectra having S/N $>1$ at each
candidate \civ\ redshift. 
The Gaussians have tied doublet separation;
free common width (constrained by FWHM$_{obs}\ge 2$ pixels); free
amplitudes $A^{1548}$ and $A^{1550}$, such that $1 \le
A^{1548}/A^{1550} \le 2$; and free velocity within $\pm 4\,000$
\kms\ of \za.\footnote {Velocity window chosen according to the
  correlation analysis explained in Sect.~\ref{sect_correlation}} Only
one \civ\ doublet is considered in each velocity window. We note that at
MUSE spectral resolution, the absorption profiles are dominated by the
unresolved LOS kinematics and not by the doublet ratios; hence,
$A^{1548}/A^{1550} \approx 1$ can occur even if the lines do not reach
zero level. This has also been observed in quasar spectra of similar
quality~\citep{Cooksey2013}.

Best-fit parameters and their 1-sigma errors were obtained for
rest-frame equivalent width ($W_0\equiv W_0^{1548} $; $\delta_{W0}$), doublet velocity
($v$; $\delta_v$), and line width ($\sigma_{\rm obs}$;
$\delta_\sigma$). We adjusted \za\ so that the median
velocity per system $\langle v \rangle=0$ \kms. The errors $\delta_v$
and $\delta_\sigma$, both relevant here, typically range within
$10$--$25$ \kms\ (68\% level; Fig.~\ref{fig_errors}). For a fit to be
considered successful (i.e., a ``detection''), we required a
significance $W_0/\delta_{W0}\ge 2$ on both doublet lines {and}
$\delta_v< 35$ \kms\ ($\approx 1/2$ pixel).  We note that the rather
loose S/N pre-selection ensured scanning of all arc spaxels, but the
final ``decision'' on detections was taken autonomously by the fitting
algorithm.  Unsuccessful fits became ``non-detections,'' for which
2-$\sigma$ upper limits were computed using $W_0=2\times$~FWHM/$\langle
S/N \rangle/(1+z)$.

\subsection{\civ\ sample}
A total of 533 binned spectra were processed, resulting in 222
detections grouped in eight \civ\ systems. We nicknamed systems with a
short name for the arcs and vowels for their incidence. 
Various maps
and properties of these systems are presented in Fig.~\ref{fig_lensed}
and Table~\ref{table_summary}, respectively.  Fitted profiles for all
systems are shown in Figs. A.1 to A.8. 
This sample comprises a unique data set of
spatially resolved and significant \civ\ detections at redshifts
$2.0$--$2.5$.

\subsection{Equivalent width completeness}
  \label{sect_completeness}

  { 
    Heterogenous S/N in a given  system naturally leads to
    an incomplete distribution of fitted parameters. 
To account for each system's $W_0$-completeness, we 
estimated the false-negative rate (FNR) of the fitting procedure. To
this end, synthetic spectra were created by replacing real absorption
features with randomly selected flux (at the continuum) from 
neighbor pixels.  
A synthetic
doublet was injected randomly inside the velocity window where the fit
was initially performed, and $W_0$ was computed as in
Sect.~\ref{sect_fitting}.  This test was conducted iteratively 
for each
binned spectrum where a feature was found
and for a certain
range of equivalent widths. Whenever the fitting algorithm failed
to detect the inserted doublet, the instance was flagged as a false
negative. 
We computed the FNR as the number of false negatives divided by the sum
of the total number of false negatives and true positives. The FNR was
computed as a function of $W_0$  and modeled 
with an exponential function of
the form $\textrm{FNR}(W_0) = \beta \exp(\alpha W_0)$.
The $W_0$-completeness level was defined as  $1 - \textrm{FNR}(W_0)$
(Fig.~\ref{fig_FNR}, upper panels) and used in Sect.~\ref{sect_correlation} 
to correct the observed $W_0$ distribution (ibid., lower
panels). Effects from S/N on other parameters are 
addressed  in Sect.~\ref{sect_direct} and
Appendix~\ref{sect_spurious_dispersion}. 
  }

\section{Spatial information and de-lensing}
\label{sect_lensing}

Clearly the most innovative -- and we assert most powerful -- feature of the
present data set is its spatial coherence and resolution. In the
so-called image plane~\citep{Grossman1988}, that is, the geometry
recorded by the instrument, the present binned spaxels are non-overlapping
squares (of 0\farcs6\ on a side); however, arcs are a consequence of
strong lensing, so the observed spaxels do not represent the real
geometry of the so-called absorber plane at
$z_{abs}$~\citep{Lopez2018}. To reconstruct the absorber plane, we used  
parametric lens models built for each field using
LenstoolTool~\citep{Jullo2007} and based on available HST imaging 
(for details, we refer the
reader to the respective publications indicated in
Table~\ref{tab:obs}). We used these models 
to calculate the
deflection angles and reconstruct the observed spaxel grid at the
absorber plane by applying the lens equation to the spaxel
vertices. The center of a spaxel in the absorber plane was taken to be
the center of a spaxel in the image plane de-lensed back to the
absorber plane.

Columns (3)--(5) of Fig.~\ref{fig_lensed} show
$2\arcsec\times2\arcsec$ stamps of the reconstructed absorber plane at
each system's redshift. Displayed are maps of $W_0$, $v$, and
absorption spread. As in column (2), these maps consist of
detections only, except for the triangles, which indicate $W_0\le 0.2$
(2-$\sigma$) upper limits. This threshold roughly corresponds to a
$50$\% $W_0$-completeness on average. 
The $W_0$ maps suggest spatial
coherence on $\sim 10$ kpc scales down to $W_0\approx 0.3$ \AA.

The figure highlights the great advantage of the present data to reach
down to sub-kiloparsec resolution at $z\approx 2$. This is due basically to
the absorber redshift being close to $z_{source}$, where beam
separations approach zero. The drawback is that in some fields, the
de-lensing produces heavily distorted grids, due to high
magnification close to critical curves. There is also some level of
overlapping spaxels, and in some cases their vertices even appear
flipped. Both effects are nonphysical and likely reflect the spatial
inaccuracy of the lens model at high magnifications. {In this
  paper, we use average separations between spaxel centers and average
  spaxel areas only; we neglect the shape of the reconstructed
  spaxels.}

The median spaxel area {of \civ\ detections} was found to be $=0.65$ kpc$^2$, so the
median linear scale is $\sim 0.81$ kpc. By system, the spaxel areas
range from $0.5$ to $3.6$ kpc$^2$. The {total} areas per system
surveyed are small compared with strong
\civ\ ``sizes'' of $\sim 100$ kpc (linear) obtained with different
methods~\citep[][]{Rauch2001,Steidel2010,Martin2010,Rudie2019,Hasan2022}.

\section{Line-of-sight versus transverse velocity dispersion}
\label{sect_direct}

\begin{figure}
  \centering
  \includegraphics[width=\columnwidth]{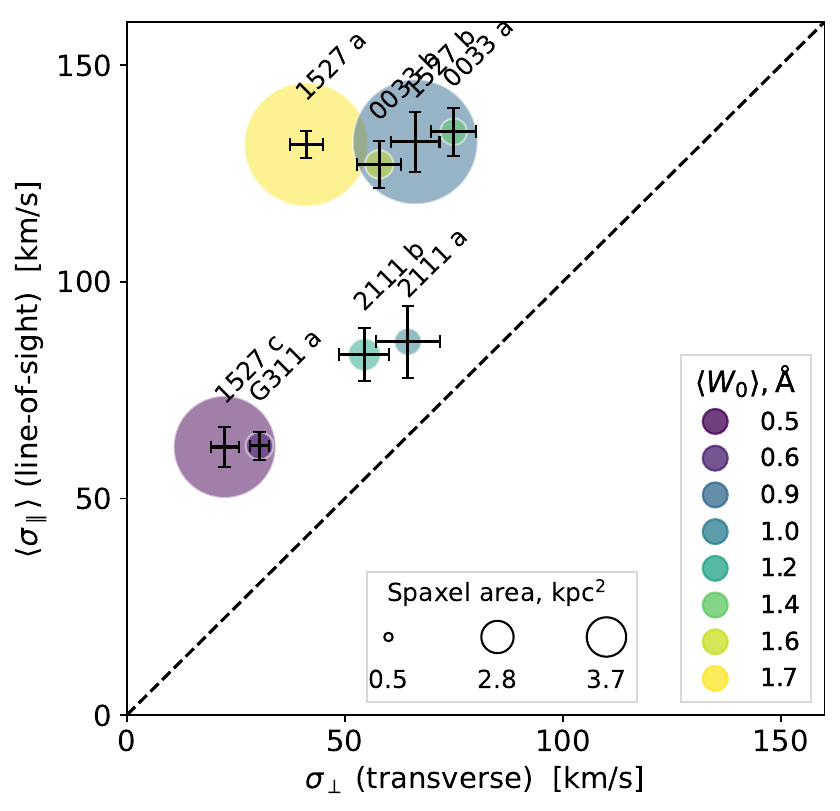}  
  \caption{System-by-system median absorption spread versus
    transverse velocity dispersion. Errors result from a bootstrapping
    analysis (Sect.~\ref{sect_direct}). 
    Circle areas are
    proportional to the absorber-plane mean spaxel area (in 
        square kiloparsecs), and 
    colors indicate 
    the median rest-frame equivalent width 
    per system (in \AA).  \spa\ has been corrected for the instrumental
    profile width (Sect.~\ref{sect_direct}), and \str\ has been corrected
    for the spurious 
    dispersion induced by each system's S/N selection function   
    (Appendix~\ref{sect_spurious_dispersion}).  
  } \label{fig_sigmas}
\end{figure}

For each system, we defined two velocity dispersions: (1) a
``transverse dispersion,'' \str, computed as the standard deviation of
$v$ of all spaxels along the arc (i.e., analogous to the projected velocity dispersion of galaxy
clusters); and (2) a ``parallel dispersion'' (also ``absorption
spread''), defined as \spa$\equiv\sqrt{\sigma_{\rm obs}^2-\sigma_{\rm
    inst}^2} $, where $\sigma_{\rm inst}=\sigma_{\rm inst}(\lambda)$
is the instrument spectral resolution (Table~\ref{table_corrections}).
  
Figure~\ref{fig_velocities} shows the distributions of $v$ and
\spa\ (blue and yellow histograms respectively). The vertical dashed
lines indicate the median $v$ and median \spa\ ($\equiv \mspa$)
Each system's 
\str\ is indicated by the length of the blue horizontal line. One
feature that stands out is that $\mspa >\str$ in all systems, despite
the fact that both dispersions vary across systems.  Also, the
velocity histograms are roughly symmetric. 
This suggests some level of
Gaussianity and little contamination by line-of-sight outliers.

The measured \str\ could be biased high due to  
velocity outliers.  On the other hand, sample incompleteness affects
the low-$W_0$ end, and since $W_0$ and \spa\ are correlated by 
the Gaussian line fitting, \mspa\ may also be
overestimated. To carry out a more robust comparison
(system-by-system) between transverse and line-of-sight dispersions, we
bootstrapped each sample by creating 1\,000 synthetic realizations of
velocities and dispersions. These are drawn randomly from ${\cal
  N}(v,\delta_v^2)$ and ${\cal N}(\spa,\delta_\sigma^2)$, where $v$,
$\spa$, and their respective variances correspond to
the measured values. 
From the bootstrapped distributions per
system, we took the median and standard deviations as corrected values.
A Monte Carlo (MC) analysis (Appendix~\ref{sect_spurious_dispersion})
showed that the S/N selection function introduces a spurious
dispersion of $\approx 17$--$29$ \kms\ (system dependent) in the spatial
direction. This value was subtracted in quadrature,
although it has only a marginal effect on \str. As for \mspa, the same 
MC analysis showed that our survey misses $\la 10$\% of
the systems detectable with MUSE (up to $\sim 30$\% in \seven\ and
\eight). It is hard to assess the effect of this
bias on \mspa, but it would translate into a  less than $5$\% only upward bias
in the ``worst-case'' scenario that all missed systems have 
$\spa\approx 0$ \kms.
  
Figure~\ref{fig_sigmas} shows the median absorption spread by system,
$\langle \spa \rangle$ as a function of \str\, with corresponding
$1\sigma$ errors.  We observed that $\mspa >\str$ remains, and we note that it
cannot be induced by the corrections {described} above.  The
relative spaxel areas probed by each system (circle sizes) do not seem
to correlate with any of the above measurements (nor the total areas
probed). Conversely, the median $W_0$ broadly correlates with $\langle
\spa \rangle$, but this is expected based on the former results from fitting
the velocity spread.

\section{Transverse auto-correlation of velocities}
\label{sect_correlation}

\begin{figure*}
  \sidecaption
    \includegraphics[width=12cm]{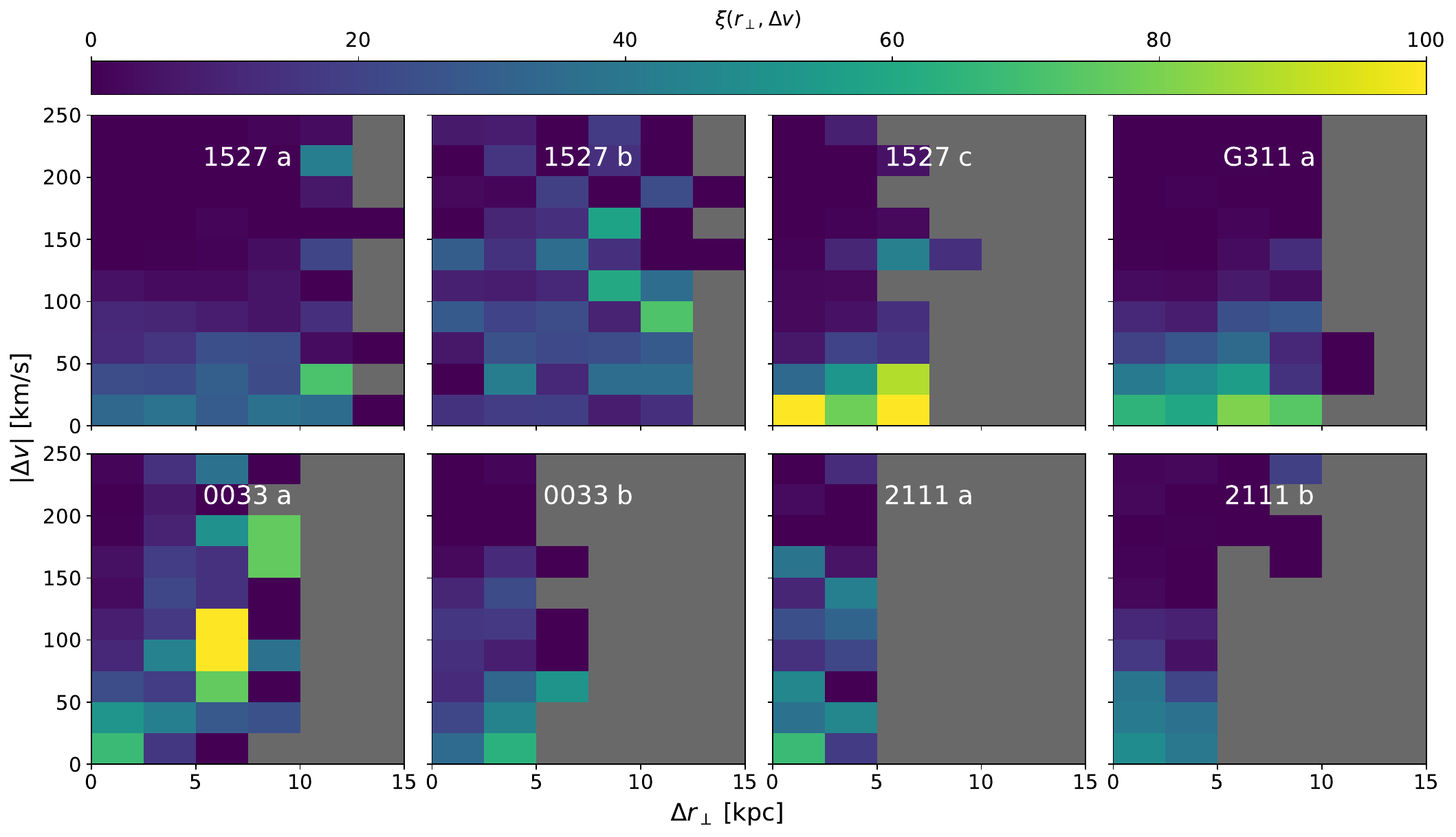}  
  \caption{\civ\ transverse velocity auto-correlation, \tpcf.   
    Gray colored bins indicate no data.
  }
  \label{fig_xi_2d_all}
\end{figure*}

\begin{figure*}
  \sidecaption
    \includegraphics[width=12cm]{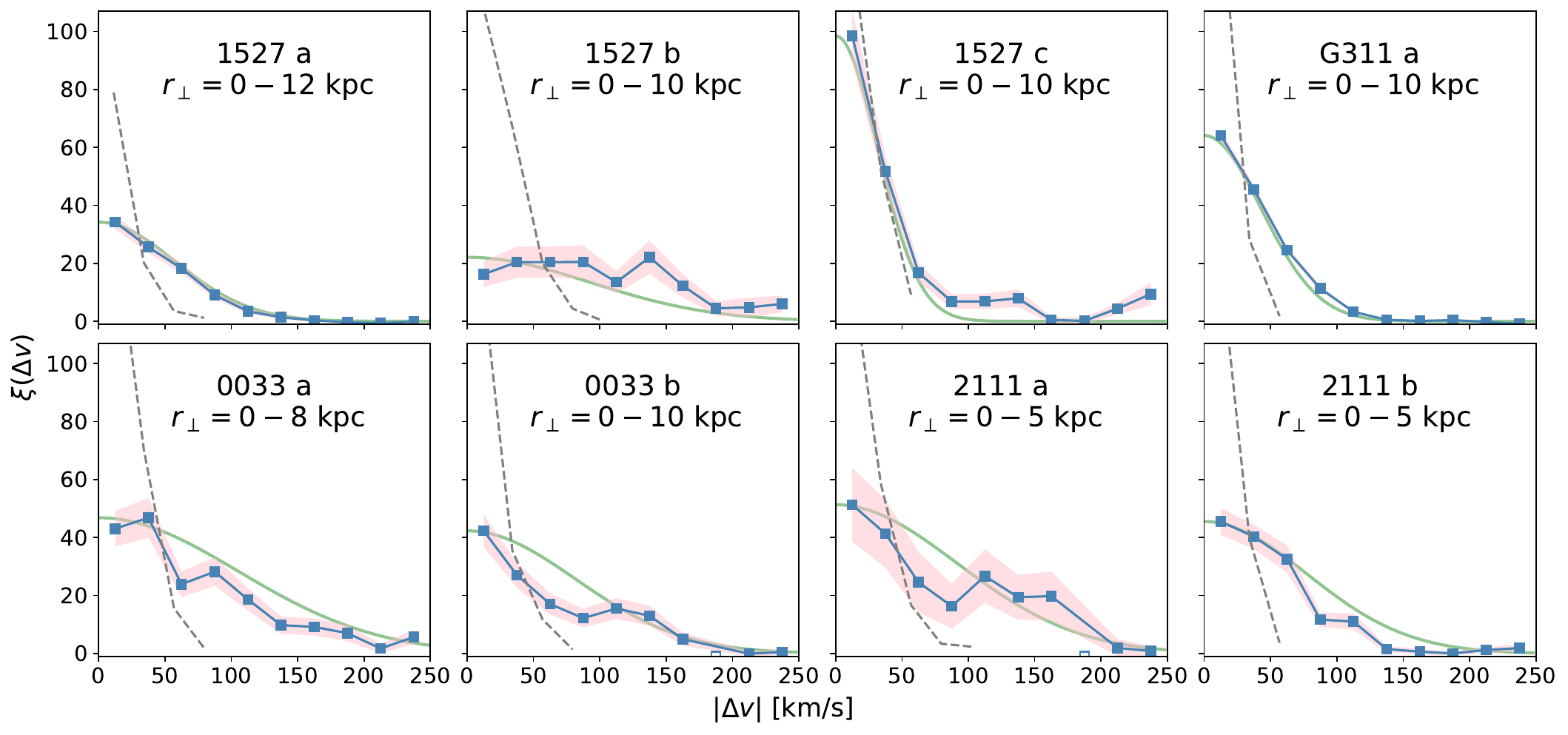}  
  \caption{
    Velocity projection of \tpcf. {Blue squares} represent the arc data. The $1\sigma$ errors (in
    pink) 
    include both the bootstrapping analysis and
    the Poisson statistics described in Sect.~\ref{sect_random} but are
    dominated by the latter. {Dashed curves} indicate the 
    signal produced by a velocity field distributed as
    ${\cal N}(0,\delta_v^2)$, where $\delta_v$ are the individual 
    measurement errors. {Green curves} show the model prediction 
    (Sect.~\ref{sect_prediction_arc}) }\label{fig_xi_1d_all}. 
\end{figure*}

Whilst the simple comparison between \spa\ and \str\ can have profound implications 
on the origin of the enriched gas and its kiloparsec-scale substructure, it is    
sensitive to outliers and hard to interpret. 
In this section, we use a more sophisticated statistical
tool by measuring coherence in pairwise velocity differences 
and separations. 
To this end, we
defined a {transverse} velocity auto-correlation function, 
computed it on a system-by-system basis, and 
tested the
possibility that this function 
(1) unveils the spatial structure and (2) is
related to the {line-of-sight} velocity correlation measured in 
quasar spectra
~\citep[e.g.,][]{Steidel1990a,Petitjean1994,Pichon2003,Scannapieco2006,Boksenberg2015,Rauch1996,Fathivavsari2013}.

\subsection{Definition}

We defined a transverse velocity auto-correlation function, \tpcf, as
the excess probability of finding a given pairwise 
\civ\ absolute velocity difference, $|\Delta v|$, and we binned the
spaxel separation on the {reconstructed absorber plane}, $\Delta r_\perp$. We used the so-called
natural estimator~\citep[e.g., ][]{Kerscher2000}:

\begin{equation}
\xi(\Delta r_\perp,\Delta v) = \frac{\langle DD \rangle}{\langle RR
  \rangle} - 1~, 
\end{equation}

\noindent where $\langle DD \rangle$ and $\langle RR \rangle$ are data-data and
random-random pair averages, respectively.
{The pairs were created from data and random catalogs described below.}
To account for errors in \tpcf, measured velocities 
were randomized with standard deviation $\delta v$, and $\langle DD
\rangle$ was computed from the median values in the $(\Delta r_\perp,\Delta
v)$ bin.  The 1-$\sigma$ errors result from adding in quadrature the
standard deviation of the median statistics (``measurement'' noise)
with the square root of the number of data pairs (``shot'' noise). {See
Appendix~\ref{sect_error_xi} for more details.}

\subsection{Data catalogs}

Each system's data catalog consists of entries of (1) a
\civ\ sight line de-lensed RA-DEC coordinate; (2) a rest-frame
equivalent width $W_0$ and its $1\sigma$ error, $\delta_W$; (3) a
velocity $v$ and its $1\sigma$ error, $\delta_v$; and (4) the same
continuum S/N level used to pre-select spaxels. Data catalogs were used to
create a list of 
pairwise velocity differences, $|\Delta
v|$, and separations on the sky, $\Delta r_\perp$.  We assumed that the
absorption signal is spatially independent from one spaxel to another.

\subsection{Random catalogs}
\label{sect_random}

Random catalogs must account for each system's S/N selection
function~\citep[e.g., ][]{Tejos2014}, both in the spectral and
transverse directions.  Here, we followed a similar procedure as
in~\citet{Martin2010} and~\citet{Mintz2022}, albeit with important
differences due to our particular data type.

For each system, we first created a new catalog filled with five repeated copies of 
the data catalog (including non-detections). This has the
advantage of preserving the survey geometry.\footnote{We did not bootstrap
because replacement would bias the pairs toward zero 
separation.} 
  Each catalog entry was populated by a
(RA-DEC$^{ran}$,$W_0^{ran},v^{ran}$)-triplet, where (1) coordinates
are uniformly distributed within the sky patch defined by all
pre-selected spaxels; (2) $W_0^{ran} $ is drawn randomly from the
completeness-corrected $W_0$ distribution, 
with replacement -- we did not use a
{model} of the $W_0$ distribution~\citep[e.g.,][]{Mintz2022}
because its shape is unknown in arc data; and (3) $v^{ran}$ is drawn
randomly from a uniform distribution of velocities, ${\cal
  U}_{[-4000,4000]}$, { 
  of spatially unresolved velocity components} in the search window,
which we consider 
the sample boundary~\citep{Mo1992}. 
(See
Appendix~\ref{sect_RRvelocities} for more details on the third condition.)  
Entries were 
rejected if $W_0^{ran} < 3 \times \delta_W$, where $\delta_W = {\rm
  FWHM}/\langle S/N \rangle/(1+z)$ is the detection limit set by the
average S/N at the wavelength corresponding to $v^{ran}$. These
entries become random non-detections. Otherwise, if accepted, the
entries become random detections. In practice, the S/N is quite flat over
the small wavelength range under consideration, implying that the
final selection is determined mostly by the spaxel-to-spaxel S/N
variations. Finally, RR pairs were built in exactly the same fashion as
DD pairs. The fidelity of the random catalogs is tested in
Appendix~\ref{sect_mock} using mock catalogs of unclustered data.

\subsection{Results}
\label{sect_results}

Figure~\ref{fig_xi_2d_all} displays \tpcf\ computed for all eight systems.
The arbitrary sampling in $\Delta r_\perp$ and $\Delta v$ minimizes the
number of no-data bins while keeping enough S/N in the less populated
bins. The signal indicates evident {velocity clustering below $\sim
  100$ \kms\ in most systems. For the systems \two\ and \five,
the velocity differences are more uniformly distributed.  
} 
In general, no spatial correlation was
observed. {If present,} either our ``resolution element''
(the total area 
per system) does not resolve it or the de-lensed coordinates are
too uncertain over these 
scales.

  Figure~\ref{fig_xi_1d_all} shows the velocity projection of \tpcf,
 {hereafter \xia,} 
created through merging bins in the spatial direction. Errors
include both the bootstrapping analysis and Poisson statistics  
but are dominated by the latter.  Consistent
with Fig.~\ref{fig_xi_2d_all}, {a significant amount of power is
 seen below $\sim 
  100$ \kms. 
For some systems (\two, \five\ and \six), a similar amount of power is
seen for velocity differences up to 200 \kms.
}
We note that
none of these signals can be mimicked by measurement errors only 
(dashed lines), except perhaps for \three. 
{Remarkably,}  
\xia\ exhibits a strong similarity 
-- in shape and amplitude -- to  the velocity two-point correlation
function measured along quasar sight lines (hereafter \xiq). 
This seems sensible because \xiq\ is measured for \civ\ "velocity components" in high-resolution (HR; $R\sim 40\,000$) 
spectra~\citep{Rauch1996,Scannapieco2006,Fathivavsari2013,Boksenberg2015}
that our data cannot resolve. This similarity is this article's
main object of study.


\section{Kinematic model} 
\label{sect_simple_model}

In this section we present a simple kinematic model that attempts to explain the results reported in Sect.~\ref{sect_direct}  
and Sect.~\ref{sect_correlation}. Subsequently, we show that this model 
provides testable clues on the kiloparsec-scale 
kinematic and spatial 
structure of  the \civ\ gas.

\subsection{Model setup}
\label{sect_model_setup}

   \begin{figure}
  \centering
  \includegraphics[trim={9cm 0cm 0cm 0cm},width=1\columnwidth,clip]{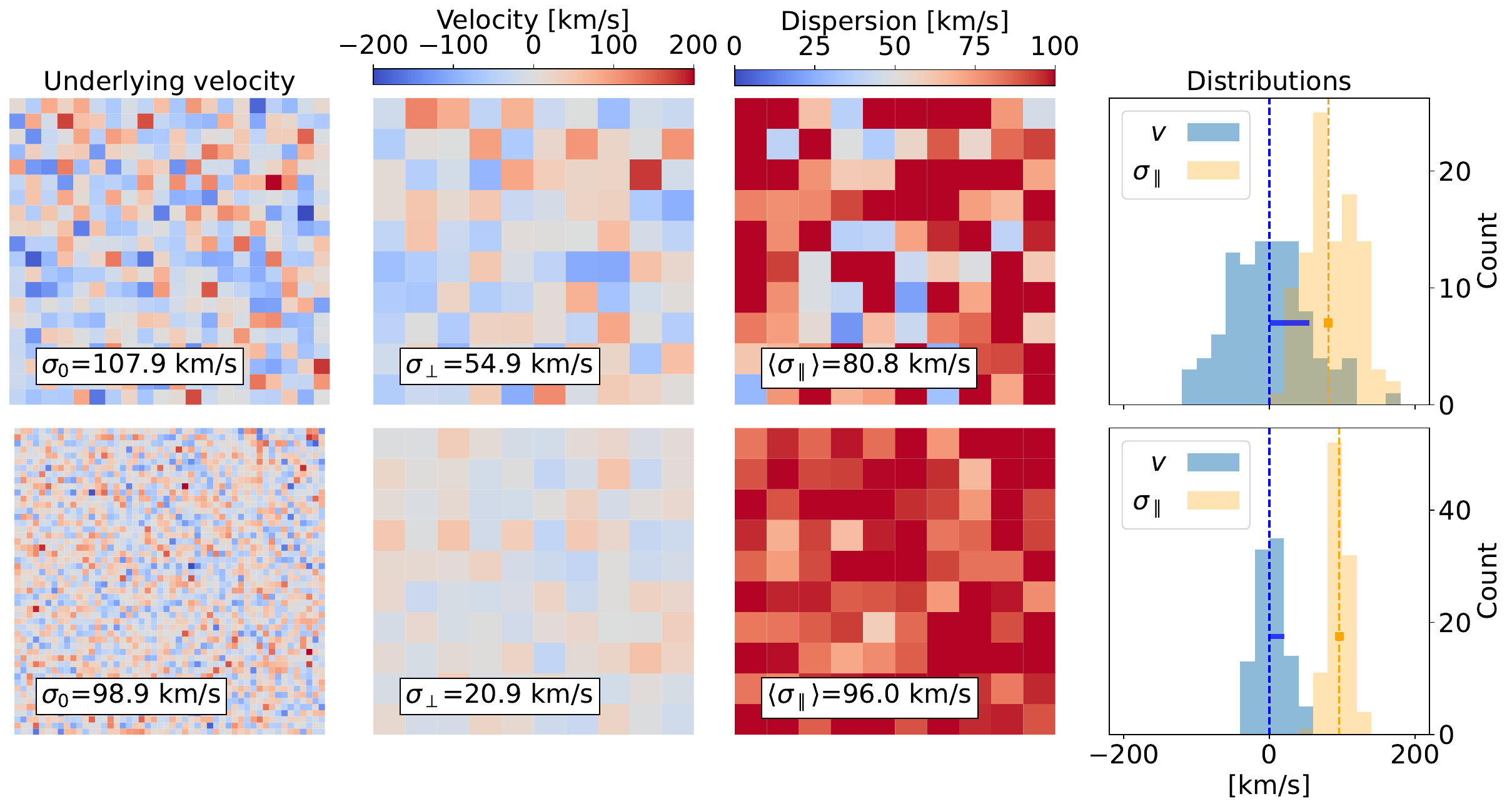}
    \caption{ 
      Kinematic model with $N=4$ (upper
      row) and $N=25$ clouds sampled per spaxel.  
      In both cases the underlying velocity field 
      distributes as 
      ${\cal N}(0,100^2)$.
      {Left-hand column}: Sampled velocities $v$ (i.e., mean of $N$ cloud
      velocities in the spaxel). Their standard deviation is the
      ``transverse dispersion,'' $\str$.  {Middle column}: Sampled
       standard deviations of the $N$ cloud
      velocities, or ``parallel'' dispersion, $\spa$.  {Right-hand column}: Sample distribution of $v$ and $\spa$. The Horizontal blue
      line represents the same as in Fig.~\ref{fig_velocities}. }
    \label{fig_sim}
  \end{figure}

We propose an underlying population of sub-spaxel absorbing gas clumps
(or ``clouds'') that the ARCTOMO observations cannot resolve spatially
but that nevertheless induce detectable absorption at the cloud velocity. 
In line with the lack of strong spatial structure seen in \tpcf, we
consider that the clouds are uniformly distributed on the sky and follow
a normal distribution of projected velocities ${\cal N}(0,\sigma_0^2)$
on the spatial scales probed here. To statistically re-create the
observations, we envisioned spaxels that sample $N$ such clouds on
average (Fig.~\ref{fig_sim}). We did not make any assumption regarding the
intrinsic position of the clouds along the line of sight.

With this simplification, we computed the statistics of the sampled
data, namely, (1) the distribution of the spaxel mean velocities and
their dispersion, $\str$, and (2) the distribution of the spaxel
velocity dispersions, $\spa$.  Examples of (1) and (2) using
$\sigma_0=100$ \kms\ and two different values of $N$ are shown in 
Fig.~\ref{fig_sim}. Computing $\str$ is exactly the same
as measuring the velocity dispersion in an ARCTOMO system. Computing
$\spa$, on the other hand, is analogous to measuring the
absorption dispersion in an ARCTOMO spectrum (but not exactly the same,
as we are not dealing with absorption profiles). These definitions
are based on the assumption that {$\spa$ conveys the kinematic
  information of individual velocity components} in the real
observations (i.e., ``within'' a spaxel).  Hence, panels (a) and (b) in
Fig.~\ref{fig_sim} are the model analog to the observations
displayed in columns 4 and 5 of Fig.~\ref{fig_lensed}, respectively.

The histograms in Fig.~\ref{fig_sim} show the distributions of $v$ and
$\spa$. They resemble the data histograms in
Fig.~\ref{fig_velocities}. Indeed, a suit of model realizations shows
that the larger the $N$, the narrower and more separated the two
distributions become. For velocities (blue histograms), this is
straightforward to see as the well-known consequence of spatial
re-sampling,  
a process which `blurs' the signal. In our
case, 
  
\begin{equation}
  \str =\sigma_0 \times \frac{1}{\sqrt{N}}~.  
\label{eq_perp}
\end{equation}

On the other hand, for $\spa$ (orange histograms), a large $N$ not
only narrows the distribution but it also shifts their peak to larger 
values. This effect is because, as more clouds are sampled per spaxel, their line-of-sight dispersions approach the original
values of $\sigma_0$. Thus, there must exist a relation also between the
original dispersion, $\sigma_0$, and the median line-of-sight
dispersion, $\langle \spa\rangle$. An analytic
solution~\citep{Kenney1951} exists for the mean only, not the median
of the $\spa$ distribution. A good (within $1\%$) approximation for
the median was found to be (Appendix~\ref{sect_model}):

\begin{equation}
  \langle \spa\rangle \approx \sigma_0 \times \frac{N-1}{N}~.   
\label{eq_long}
\end{equation}

As expected, the larger the number of clouds in a spaxel, the more
the recovered dispersion approaches the original one. 
We note that the two equations above hold true independent of the physical spaxel size and depend only on $N$ and $\sigma_0$.

\subsection{Model predictions}
\label{sect_predictions}

\subsubsection{Transverse versus parallel velocity dispersion}

\begin{figure}
  \centering
  \includegraphics[width=1\columnwidth]{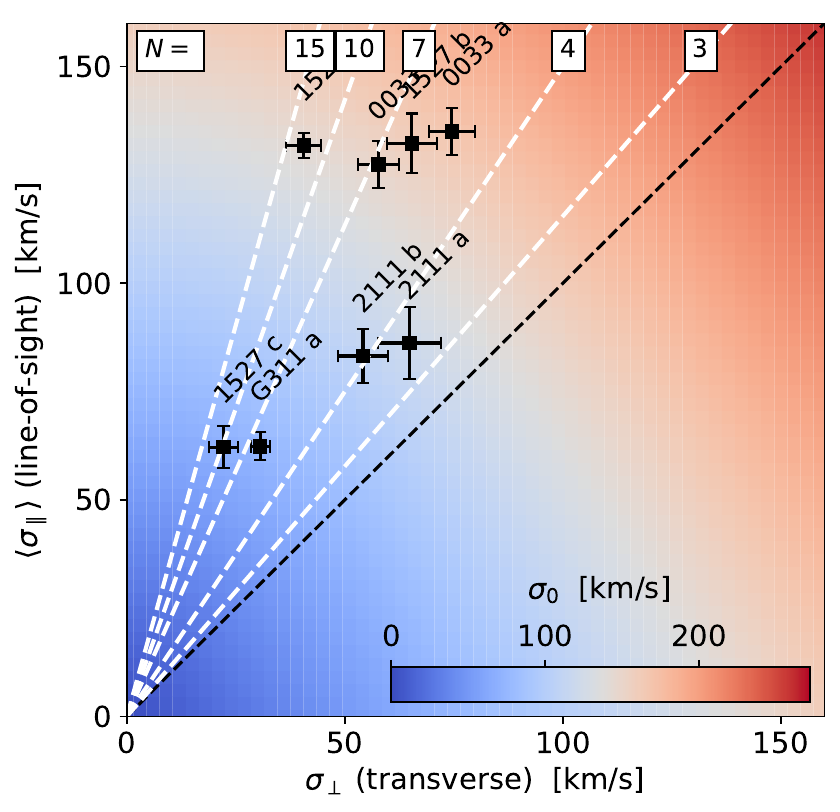}  
  \caption{Same as Fig.~\ref{fig_sigmas} but showing the
    two predictions of our kinematic model on (a) $\langle
    \spa\rangle$ versus \str\ for a given number of
    clouds per spaxel, $N$ (white dashed lines;
    Eq.~\eqref{eq_N}), and (b) the underlying line-of-sight
    dispersion, 
    $\sigma_0$ (background color; Eq.~\eqref{eq_sigmas}).
  } \label{fig_sigmas_prediction}
\end{figure}

Equations~\eqref{eq_perp} and~\eqref{eq_long} show that the spatial
sampling determines $\str$ and $\langle\spa\rangle$ and that this is
a purely observational effect. Since we can measure $\str$ and
$\langle\spa\rangle$ in ARCTOMO data, the equations can be used to
derive two intrinsic properties of the cloud distribution, $N$ and
$\sigma_0$.  Solving for $\sigma_0$, one obtains:

\begin{equation}
  \langle \spa\rangle =\str \times  \frac{N-1}{\sqrt{N}}~,   
\label{eq_N}
\end{equation}

from which we derive $N$, the per-spaxel mean number of clouds.  In
Table~\ref{table_sigmas}, $N$
is listed for each system. Equation~\eqref{eq_N} can also be used to
re-create Fig.~\ref{fig_sigmas} for particular values of $N$. This is
shown in Fig.~\ref{fig_sigmas_prediction}, where the white dashed
lines display the $\langle \spa \rangle$ versus \str\ relation for
selected values of $N$.  Since $N$ does not need to be an integer, we
interpreted it as the average number of clouds per spaxel. Equation~\eqref{eq_N} also explains why, under the assumption of
Gaussianity, one should generally measure $\langle \spa \rangle
>\str$ as a result of the particular ARCTOMO observations
``flattening'' the underlying velocity field.\footnote{Interestingly,
  $\mspa=\str$ for $N=1+\phi$, where $\phi$ is the ``Golden ratio.''}

Likewise, arranging Eqs.~\ref{eq_perp} and~\ref{eq_long}, this
time to get rid of $N$, one finds

\begin{equation}
  \sigma_0 = \frac{\langle \spa\rangle \pm
  \sqrt{\langle\spa\rangle^2 +4\str^2}}{2}~.
\label{eq_sigmas}
\end{equation}

Finally, 
an equation
emerges 
that gives the underlying \civ\ velocity
dispersion out of two ARCTOMO observables.  The solutions for each
system are listed in Table~\ref{table_sigmas}, and the background of
Fig.~\ref{fig_sigmas_prediction} displays $\sigma_0$ computed for all
({\str},$\langle \spa \rangle)$ combinations, as indicated by the
color scale. The uncertainties in Table~\ref{table_sigmas} were
propagated from those in \spa\ and \str. 
As a sanity check for
Eq.~\eqref{eq_sigmas}, if \str$=0$ (no transverse dispersion), then
$\sigma_0 = \langle \spa\rangle$; that is, the underlying dispersion is
recovered (and all spectra show the same centroid velocity). The
second solution, $\sigma_0 =0$ (no absorption), is physical only for
\spa$=0$.

  \begin{table}
\centering
\begin{threeparttable}
\caption{Model predictions.}
\begin{tabular}{lcccc}
\hline
\hline
System&$N$&$\sigma_0$&$\langle N_c\rangle$&$S_c$\\
&&[km/s]&[kpc$^{-2}$]&[kpc]
\\
&(1)&(2)&(3)&(4)\\\hline
1527 a&12.4$\pm$3.4&143.0$\pm$4.9&3.4$\pm$1.1&0.5--1.9\\
1527 b&5.9$\pm$1.7&159.8$\pm$8.9&1.6$\pm$0.5&0.8--1.9\\
1527 c&9.5$\pm$4.7&69.0$\pm$5.6&3.3$\pm$1.8&0.5--1.7\\
G311 a&6.0$\pm$1.4&74.4$\pm$3.7&12.0$\pm$3.5&0.3--0.7\\
0033 a&5.1$\pm$1.2&167.9$\pm$7.6&10.6$\pm$3.0&0.3--0.7\\
0033 b&6.7$\pm$1.8&149.6$\pm$7.4&13.0$\pm$4.2&0.3--0.7\\
2111 a&3.5$\pm$1.5&121.2$\pm$11.1&7.1$\pm$3.3&0.4--0.7\\
2111 b&4.1$\pm$1.5&110.2$\pm$8.5&6.7$\pm$2.8&0.4--0.8\\
\hline
\label{table_sigmas} 
\end{tabular}
\begin{tablenotes}[flushleft] 
\item Table columns: 
(1) number of clouds per spaxel, 
(2) underlying velocity dispersion, 
(3) mean number of clouds per unit area, and
(4) cloud sizes for covering fraction $\kappa=1$ (Sect.~\ref{sect_sizes}). 
Errors are propagated from \str\ and \spa\ uncertainties. 
\end{tablenotes}
\end{threeparttable}
\end{table}

  \subsubsection{Inter-cloud distances}  
  \label{sect_intercloud}

So far, 
our model does not involve physical scales, but we can define the
(projected) mean number of clouds per unit area as $\langle
N_c\rangle\equiv N/A_{spaxel}$, where $A_{spaxel}$ is the mean spaxel
area per system~\citep[hence, $\langle N_c\rangle$ is equivalent to
  the ``counts-in-cylinder'' statistics on galaxy
  scales;][]{Berrier2011}.  With this definition, the projected mean
distance between clouds is

\begin{equation}
    \dctwo \sim \left(\frac{1}{\langle N_c \rangle}\right)^{1/2} {\rm [kpc].}
    \label{eq_dc2}
  \end{equation}

Using the values for $\langle N_c\rangle$ listed in
Table~\ref{table_sigmas},   
the present systems occur in
structures separated on the sky by  $\dctwo = 0.3$--$0.8$ kpc. We
discuss the implications of \dctwo\ on cloud "sizes" in Sect.~\ref{sect_sizes}. 

Likewise, $\langle N_c\rangle$ could constrain the number density of
clouds $n_c = N_c/L$, where $L$ [kpc] is the (unknown) total
absorption length. In this case, the mean 3D distance between clouds
is

  \begin{equation}
    \dcthree \sim \frac{1}{n_c^{1/3}} \sim
    \left(\frac{L}{\langle N_c \rangle}\right)^{1/3} {\rm [kpc].}
    \label{eq_dc3}
  \end{equation}

From Table~\ref{table_sigmas}, we observed that the present systems arise in
structures separated in space by $ \dcthree = 2$--$4~(L/100)^{1/3}$
kpc. Current estimates on the extension of $W_0\ga 0.3$
\AA\ \civ\ halos amount to $\approx 100$--$200$
kpc~\citep{Steidel2010, Hasan2022}.

\subsubsection{Cloud sizes and covering fraction}
\label{sect_sizes}
  
{While our model does not constrain cloud ``sizes'' directly, our
  results suggest that cloud sizes cannot exceed the arc beams.  In
  fact, if the clouds were much larger than the beam, we would measure
  (within observational uncertainties) the same velocity centroids
  across the entire arc and therefore also $\str\approx 0$.  As a
  result, \xia\ would be much steeper (given that $\Delta v\approx 0$
  across spaxels) than what we measure (Fig.~\ref{fig_xi_1d_all}).
  The only way to produce the observed signals at larger $\Delta v$ is by
  having a cloud size that is about the same size or smaller than the
  spatial scale probed by a spaxel.

On the other hand, our model predicts a projected inter-cloud
distance, \dctwo\ (Eq.~\eqref{eq_dc2}). This parameter is related  to
the cloud characteristic size, $S_c$, and to 
the covering fraction, $\kappa$, defined  as the
fraction of the beam area covered by all clouds in projection along the
line of sight.
Our data cannot directly disentangle $S_c$ and  $\kappa$, but the
following scenarios seem plausible:

\begin{enumerate}[label={[\arabic*]}]
\item 
$\kappa\approx 1$, at the limit of
no overlap: This implies cloud sizes of $S_c\approx \dctwo$. 
\item  
$\kappa\approx 1$, with
considerable overlap: This implies $S_c\ga\dctwo$ but below the beam
size for 
the reasons outlined above. 
\item  
$\kappa< 1$:  In this 
case, sizes remain unconstrained but below $\dctwo$.
\end{enumerate}

Table~\ref{table_sigmas} displays 
$S_c=\dctwo$ (first scenario, which we consider the most likely), using Eq.~\ref{eq_dc2} and the tabulated $\langle N_c
\rangle$ values.
} 
We emphasize that these considerations apply to clouds producing
detectable ARCTOMO signals (i.e., typically $W_0\ga 0.3$ \AA).

In
conclusion, our kinematic model constrains strong \civ\ cloud sizes to
be of the order of or smaller than the probed beam size (i.e., $\la 1$
kpc). 
{ 
Future higher spatial and spectral resolution data should
    help discern between the  
scenarios described above.
}

\subsubsection{Transverse auto-correlation}
\label{sect_prediction_arc}

Our kinematic model can also predict  \xia\  
(Sect.~\ref{sect_results}). Recalling that if a
random variable $X$ is normally distributed with variance $\sigma^2$, then
$\Delta X$ is normally distributed, too (with variance $2\sigma^2$), 
Eq.~\eqref{eq_perp} implies that

\begin{equation}
\sigma_\xi^{arc} = \sigma_0 \times \sqrt{\frac{2}{N}}~,
\label{eq_xi}
\end{equation}
where $\sigma_\xi^{arc}$ is the $1\sigma$ width of \xia.  The green
curves in Fig.~\ref{fig_xi_1d_all} compare this prediction with the
measured \xia. Since the model only predicts the width of \xia, we
display a Gaussian with an amplitude given by the maximum value of
\xia\ and a width given by Eq.~\eqref{eq_xi}, using $\sigma_0$ and $N$
listed in Table~\ref{table_sigmas} (a comparison between {fitted}
and predicted widths is shown in Fig.~\ref{fig_xis}). The measurement and
the prediction match {reasonably} well, which demonstrates the
self-consistency of the model (since both quantities are based on the
same data).

\subsubsection{Prediction using the quasar line-of-sight correlation}
\label{sect_prediction_quasar}

\begin{figure}
  \centering
  \includegraphics[width=\columnwidth]{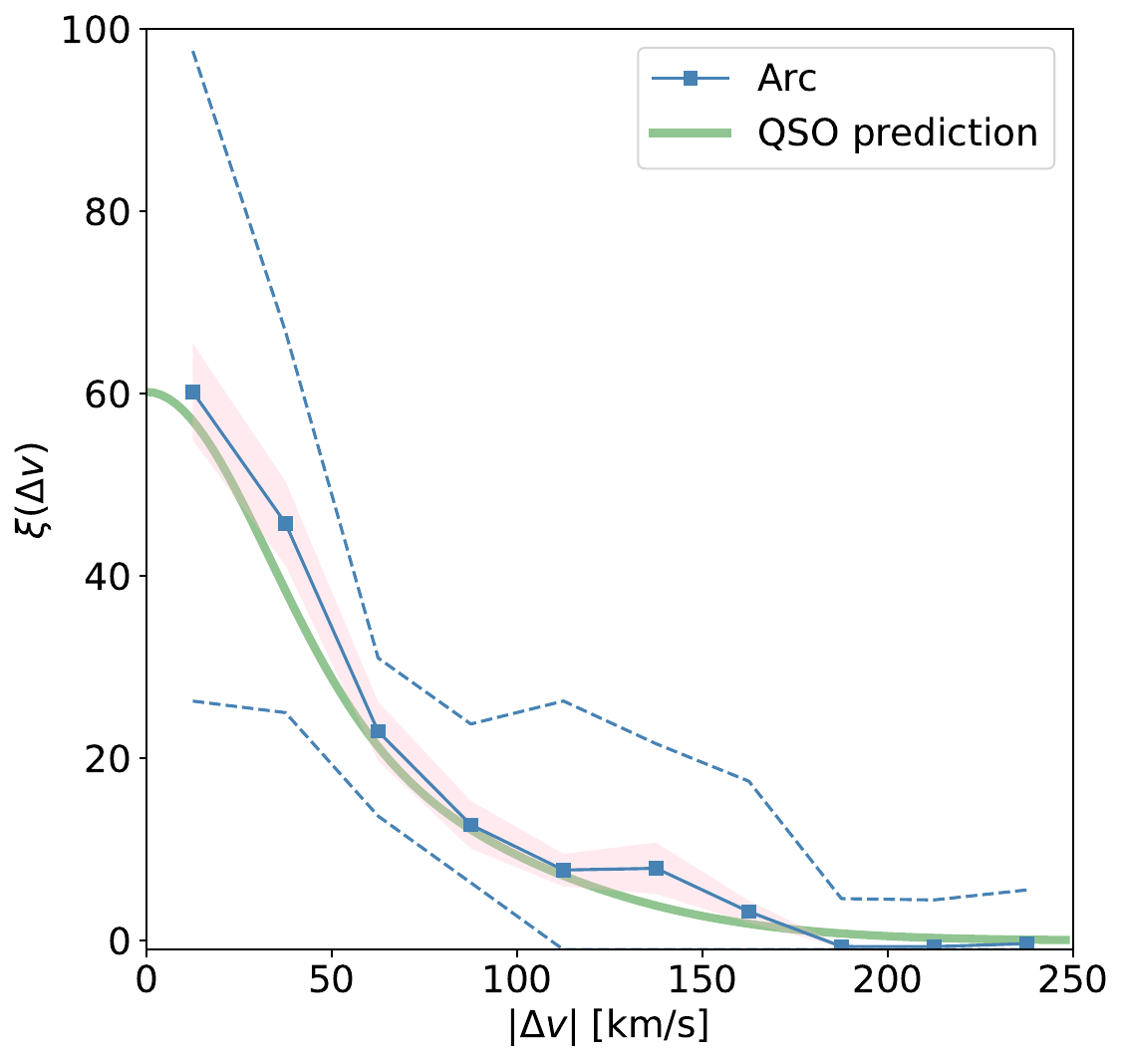} 
  \caption{
    Stacked average \xia\ of all eight arc systems 
    and QSO prediction (details in~Sect.~\ref{sect_prediction_quasar}).  
    The dashed curves indicate
        the range of arc systems. 
  } \label{fig_xi_ave}
\end{figure}

As an independent test, our model can also predict
\xia\ from 
\xiq.  To this end, we
first stacked and averaged \xia\ over all eight systems. 
Next,  we assumed that the \civ\ clouds abide to a normal
distribution that is captured by both \xia\ and \xiq; in other words, {the
clouds are responsible for both the parallel signal toward quasars and
the transverse signal toward arcs.} 
{This is a strong assumption that we discuss 
  further below 
in Sect.~\ref{sect_structure}.   
From Eq.~\eqref{eq_xi}, it follows that 
$\sigma_\xi^{arc}=\sigma_\xi^{quasar}/\sqrt{N}$.} 
{\citet{Boksenberg2015} fit
\xiq\ with the sum of two Gaussians having  $\sigma_\xi^{quasar}=80$ and $185$
\kms\ (also with a narrower one that our data do not
resolve).} \citet{Scannapieco2006} and \citet{Fathivavsari2013} do not 
provide fits, but their correlation functions have widths consistent
with these. In Fig.~\ref{fig_xi_ave}, we display a 
comparison between our transverse correlation and the line-of-sight  
``QSO prediction'' using $N=6.7$, 
the weighted 
average number of clouds per spaxel in the arc data. 
The match is quite good 
($\lesssim 2\sigma$ deviation), 
lending {independent} support to our  kinematic model and to the inferred
number of clouds per spaxel. 
It also suggests that the present (modest) number of \civ\ systems is only
moderately affected by cosmic variance.

\section{Discussion}
\label{sect_discussion}

  \begin{figure*}
  \centering
  \includegraphics[width=2\columnwidth,clip]{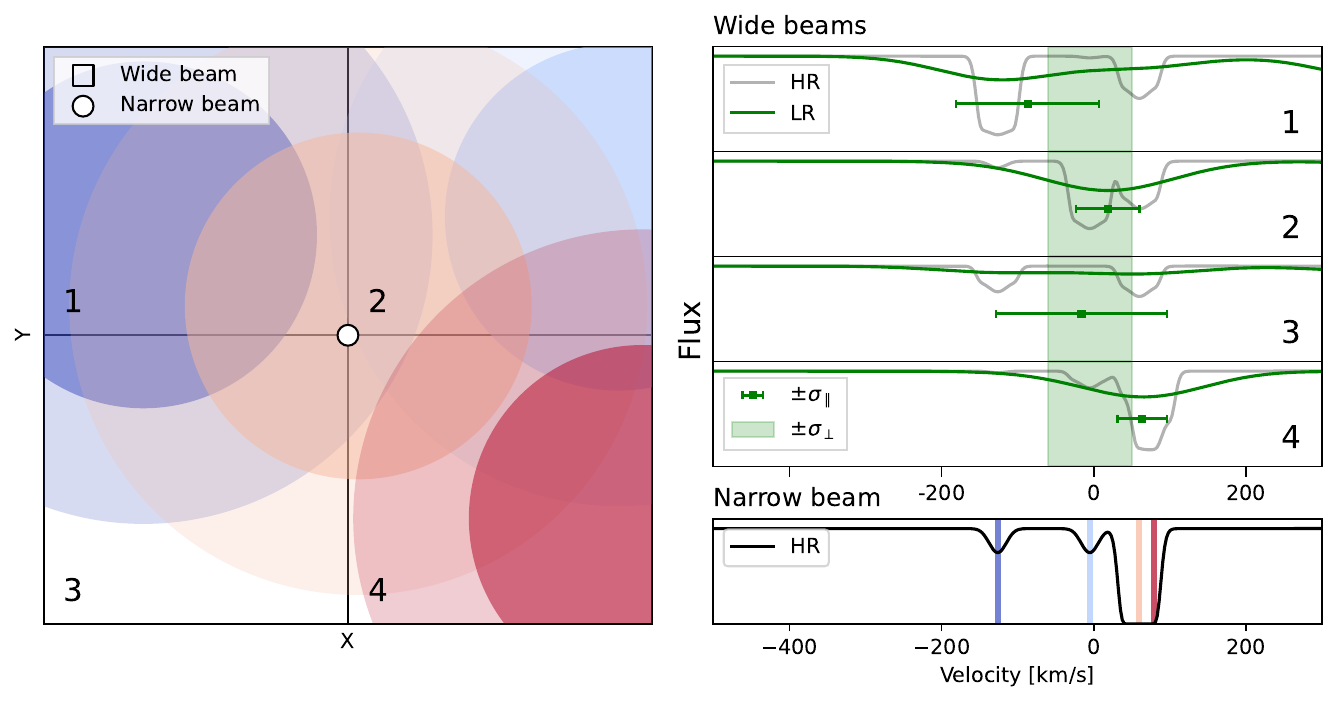}  
  \caption{ Semi-schematic visualization of the kinematic model. 
      Left-hand panel: Column density map, $N({\bf  x}$). 
    Here, ${\bf  x}$ denotes the on-sky position. The map was created by superposing four clouds
    with $N(\civ)=10^{14}$ \icm\ (cores) and $10^{12}$
    \icm\ (halos) at random positions and line-of-sight velocities,
    $v_c$. The clouds are colored according to $v_c$.  The white dot
    represents a narrow beam (e.g., a quasar pencil beam), and the four
    squares indicate wide beams (e.g., ARCTOMO spaxels). An optical-depth cube
    $\tau(v,{\bf  x})$ was created by evaluating Voigt
    profiles~\citep{Liang2017} with $v=v_c$, $N=N({\bf  x})$, and
    $b=10$ \kms, and spectra were generated by averaging $e^{-\tau}$
    within the beam area. {Upper right panels:} Wide-beam spectra
    at infinite and MUSE spectral resolution (HR and LR,
    respectively).  The green squared symbols with error bars indicate
    LR individual line velocity centroids and \spa\ dispersions,
    respectively, obtained from Gaussian fits. The shaded region
    indicates the overall \str\ dispersion of the four LR velocities.
    { Lower right panel:} Narrow-beam spectrum at infinite spectral
    resolution. We note that $v_c$ is indicated with the same colors as in the
    left-hand panel.  } \label{fig_cartoon}
  \end{figure*}

  The proposed kinematic model seems acceptable on two {different}
  fronts: a self-consistent comparison between ARCTOMO transverse and
  line-of-sight kinematics, and a prediction of the transverse velocity
    auto-correlation from the quasar line-of-sight kinematics. The
  first one is solely observational, as we have outlined above. But
  the second suggests that {the weak \civ\ components (only
    detected in HR quasar spectra) and the strong components (our arc
    detections) may
    trace the same physical structures.} 
  We conclude this article by discussing how our results are
    consistent with even more  quasar observations and how together they
    provide glimpses of the 
CGM on both the sub-kiloparsec and galactic scales.

    \subsection{Cloud density structure}
\label{sect_structure}
{ 
A natural explanation for a single cloud population giving rise
to both strong and weak components is that each cloud has an internal
density structure. 
Consequently, as a second-order approximation of the model, in the
following we refer to cloud ``cores'' (the central parts of the clouds
responsible for the strong components) and ``halos'' (external parts of
the clouds responsible for the weak components).
  
Figure~\ref{fig_cartoon} illustrates the proposed situation.  The
left-hand panel displays a column-density ``map'' of cloud cores and
halos. The white dot represents a narrow beam (e.g., a quasar beam),
and the squares represent the arc beams (e.g., ARCTOMO spaxels). In this simple
example, the beams  
pierce a clumpy medium represented by the superposition of just four
clouds. The right-hand panels display
the respective spectra. These  
result from spatially averaging the flux within
the beam areas (see the figure caption for more details on the
implementation). By model construction, the narrow beam collects the
kinematic information of many clouds through intercepting mostly their
larger cross-section cloud halos. Each cloud produces a velocity
component in an HR spectrum (lower-right panel). Conversely, a 
LR spectrum of 
the wide beam recovers only an average velocity and a line-of-sight
dispersion (\spa), and 
it is dominated by a few cloud cores inside the beam
(upper-right panels). {As expected, 
  $\langle \spa \rangle >\str$
  (see the 
 figure caption for how \spa\  and \spa\  are represented). }

{A scenario of clouds producing both weak and strong components
  also} 
explains why
removing the high column-density components from quasar HR data
does not affect \xiq~\citep{Scannapieco2006}. Those authors find
little change in \xiq\ when $N=10^{13,14,15}$ \icm\ components are
excluded (up to 10\% of their sample, according to the reported column density frequency). In the present  
interpretation, this effect is not
simply due to the strong components being a small fraction of the
total but more fundamentally to
both strong and weak {components} sharing the same velocity field.

A cloud density structure equivalent to our model has already been
proposed. 
For instance, 
\cite{Hummels2024} have introduced the concept of a cool CGM 
     {complex,} {which is composed of multiple cloudlets
of various masses and velocities} that lead to a column density
structure, as in our 
case. 
However, we emphasize that our model aims to explain the observed
kinematics only. 
Besides, the canonical one-to-one 
  association between the 
  cloud and the velocity component 
    may also be  too
simplistic~\citep[e.g.,][]{Faerman2023,Marra2024, Li2024}.

  \subsection{Other multiple sight line observations}

The most direct comparison between arc and quasar results can be made
with the --- unfortunately very limited --- sample of multi-sight line
quasar observations 
subtending approximately kiloparsec separations. The first direct estimate of
\civ\ coherence length using lensed quasars~\citep{Rauch2001a} delivered
$\sim 0.3$ kpc (50\% transverse variations). This value is consistent
with subsequent studies that found \civ\ transverse structure below 1
kpc~\citep{Tzanavaris2003,Lopez2007,Rubin2018}. Along with the
ten times larger \dcthree\ computed in Sect.~\ref{sect_intercloud}, the
above narrow-beam size constraints imply a small volume filling factor
of $\sim 10^{-3}$, which is in turn comparable with theoretical
predictions~\citep{McCourt2018,Gronke2020,Liang2020,Li2024}. On the
other hand, combining those sizes with our \dctwo\ constraint
(Eq.~\eqref{eq_dc2}) leads to a near-unity covering fraction. This is
again consistent with theoretical predictions~\citep{Liang2020}. We
conclude that our arc results are consistent with quasar
observations and 
theoretical predictions of the \civ\ spatial domain.

Regarding the line-of-sight direction, we measured $\mspa\approx
60$--$130$ \kms, a range which seems consistent with the velocity
shear found toward lensed quasars. \citet{Rubin2018} find a velocity
structure $\Delta v\approx100$ km/s over kiloparsec scales;
\citet{Rauch2001a} found $\approx 60$ \kms\ over $10$ kpc;
\citet{Ellison2004} measured $\approx 15$ \kms\ on sub-kiloparsec scales; and
\citet{Lopez2007} found some cases with $60$ \kms\ shear over $1$
kpc. Thus, also in the parallel direction, {arc} and quasar observations
seem consistent with each other.

\subsection{Departures from Gaussianity and large-scale kinematics}

Our model assumes an underlying Gaussian velocity field. In order to
gauge departures from this idealization, we considered a spatial gradient
in velocity, $\Delta v$, produced, for instance, by galaxy-scale
motions such as orbiting, co-rotating, or out-flowing gas. Assuming the
spatial scale of $\Delta v$ is much larger than the spaxel size, it is
straightforward to see that $\str$ will increase but $\spa$ will not. This
is because the former is an inter-beam measurement, whereas the latter
is an ``intra-beam'' one. In this case, Eq.~\eqref{eq_perp} does not
hold anymore, and in Fig.~\ref{fig_sigmas} all points become shifted to
the right. A simple numerical test shows that $N$ becomes
underpredicted as soon as $\Delta v$ is comparable to $\sigma_0$.
Thus, departures from Gaussianity could be detected in ARCTOMO systems
exhibiting $\str \ga \langle \spa \rangle$ and \tpcf\ power both at
large spatial and velocity scales. None of the present systems clearly
qualifies in this group, {but  
in principle our experiment could be used 
to identify large-scale kinematic motions of the absorbing CGM. }

{On the other hand, the large underlying velocity dispersions we
  derived} 
are intriguing. The largest $\sigma_0$ values
in Table~\ref{table_sigmas}  
suggest
velocity dispersions of galaxy groups, and one possibility is that the
unresolved clumps are bound to different group galaxies. Another one,
of course, is that they feature the long-sought manifestation of
super winds~\citep{Voit1996,Pettini2001}. Associations between
\civ\ in quasar spectra and star-forming galaxies find kinematic
separations at this level, and these have been associated with
clustering~\citep{Adelberger2005,Lofthouse2023,Banerjee2023,Galbiati2023},
inflows~\citep{Turner2017}, or filaments outside the galaxy virial
radius~\citep{Galbiati2023,Banerjee2023}.  All of these tests suggest
that \civ\ in quasar spectra trace galaxies, although none of them
really can disentangle the absorbing galaxy from the absorption
system. Our transverse observations literally add a new dimension to
the understanding of the origin of metal-rich gas in the $z=2$--$3$ 
cool-warm CGM.

\subsection{The path forward}

It {may} seem surprising that the present ARCTOMO data, although
unable to 
resolve the absorbing clouds neither spatially nor spectroscopically,
still carry the 3D kinematic information encoded in their absorption
profiles. We have shown that {these data, along with
  simple assumptions about the kinematics, can lead to a set of
  realistic and testable predictions.}

But the current tomographic data do not yet allow for a complete
disentangling between different global dynamic models of the
\civ-bearing CGM. The definitive pieces of the puzzle will be obtained
through (1) the detection of the \civ\ galaxies responsible for the
arc signals, a challenging objective that requires space-based near-infrared
observations, and (2) spatially resolved HR observations of extended
background sources, where little has been done
yet~\citep[e.g.,][]{Diamond-Stanic2016}. Along with more
sophisticated models that include density structure, we assert that
such observations shall dramatically improve our understanding of the 
cool-warm CGM's small-scale structure. We consider, for instance, the HR
curves in the upper-right panels of Fig.~\ref{fig_cartoon}. If the
kinematic model tested in this article is reliable, a small number of clouds per 
spatial resolution element of approximately kiloparsec size should unfold as
resolved velocity components.

\section{Summary}
\label{sect_summary}

We have presented spatially resolved VLT/MUSE observations of
four giant gravitational arcs that offer \civ\ coverage. We focused on the kinematic properties of \civ\ absorption detected
at $\langle z_{abs}\rangle \sim 2.3 $, a redshift that enables sub-kiloparsec resolution of
the absorbers thanks to lens magnification. Our experimental setup allowed us to analyze \civ\ absorption in 222 adjacent,
  uncorrelated  beams 
that pierce eight intervening
\civ\ systems. Our results are as follows:

\begin{enumerate}
\item  
Significant absorption is detected across almost all the arcs,
implying \civ\ extensions ($W_0\ga0.3$ \AA) of at least $\approx 10$
kpc in the reconstructed absorber plane (Fig.~\ref{fig_lensed}). We
ran an automated search of \civ\ absorption and computed doublet
velocities, absorption spreads, and equivalent widths. Each system
shows evident velocity clustering in the transverse direction
(Fig.~\ref{fig_velocities}), and we set out to measure such clustering
and investigated its origin.

\item
On average, the transverse velocity dispersion, \str, is found to be
smaller than the per-system median line-of-sight dispersion, $\langle \spa
\rangle$ (Fig.~\ref{fig_sigmas}). 

\item 
  To measure spatial clustering, we
  computed a transverse auto-correlation function of
\civ\ velocities (\tpcf; Fig.~\ref{fig_xi_2d_all}) and its velocity
projection (\xia; Fig.~\ref{fig_xi_1d_all}).  \tpcf\ does not show
evident spatial patterns, perhaps due to insufficient resolution. On the
other hand, 
the average \xia\ exhibits great
similarity in shape and amplitude with the line-of-sight two-point
correlation measured in HR quasar spectra, \xiq.

\item
To aid a comparison between wide- (e.g., arc) and narrow- (e.g., quasar)
beam observations, we introduced a simple kinematic model in
which the absorption profiles result from groups of $N$ clumps (clouds) sampled 
at the sub-spaxel level and that produce a mean velocity and
line-of-sight dispersion (Fig.~\ref{fig_sim}). The model successfully
explains the $\str< \spa$ inequality under the assumption of an
underlying Gaussian field with dispersion $\sigma_0$. It also
consistently predicts $N$ (Eq.~\eqref{eq_N}) and $\sigma_0$
(Eq.~\eqref{eq_sigmas}) out of the observables \str\ and
\spa\ (Fig.~\ref{fig_sigmas_prediction}). Combining this information
with the average spaxel areas allowed us to constrain the number of
clouds per unit area to within $N_c\approx 2$--$13$ kpc$^{-2}$ and
$\sigma_0\approx 70$--$170$ \kms, depending on the system. The model also
constrains the projected mean inter-cloud distance to within $\approx
0.3$–-$0.8$ kpc. The covering fraction and kinematic considerations put
strong \civ\ sizes between those values and the spaxel size,
$0.7$--$1.9$ kpc.

\item 
Our model also predicts the width of \xia\ out of the independently
computed \xiq\ reasonably well. Since \xiq\ is dominated by the bulk
of weak velocity components that the arc data cannot resolve, this
match led us to conclude that a single population of \civ\ 
  clouds must be responsible for 
both signals. This in turn implies that the clouds must have a
density structure, for instance, a radial density profile that
produces a few strong components that shape \xia and many weak
components that shape \xiq\ and only get detected in HR spectra (e.g.,
of quasars; Fig.~\ref{fig_cartoon}). In such a scenario, \xiq\ is
dominated by the weak components, as has been noted elsewhere.

\item
  We have discussed how our model and observations are compatible with
extant observations of multiple quasars, both in the transverse and the
parallel dimensions.

\end{enumerate}

Upcoming large optical facilities will routinely enable tomography of
the CGM using not only lensed galaxies but also normal
galaxies.  Comparison of these wide-beam observations with the 
extant narrow-beam statistics, as shown here for a handful of systems,
promises a better understanding of the connection between the small-
and the galactic-scale structure of the high-redshift CGM.



\begin{acknowledgements}

We would like to thank the anonymous referee for a thorough and
critical review of our analysis. This work has benefited from
discussions with Max Gronke, Joseph Hennawi, and Claudio Lopez.
S. L. and N. T. acknowledge support by FONDECYT grant
1231187. M. S. was financially supported by Becas-ANID scholarship
\#21221511, and also acknowledges ANID BASAL project FB210003.

\end{acknowledgements}

\bibliographystyle{aa}
\bibliography{civ_lit}

\begin{appendix}

\section{Absorption line profiles and fit results}

Individual absorption systems are displayed in
Figures~\ref{fig_data_1} to~\ref{fig_data_8}. The corresponding
best-fit parameters are listed in Table~\ref{table_bestfit}.

\begin{figure}
  \includegraphics[width=\columnwidth]{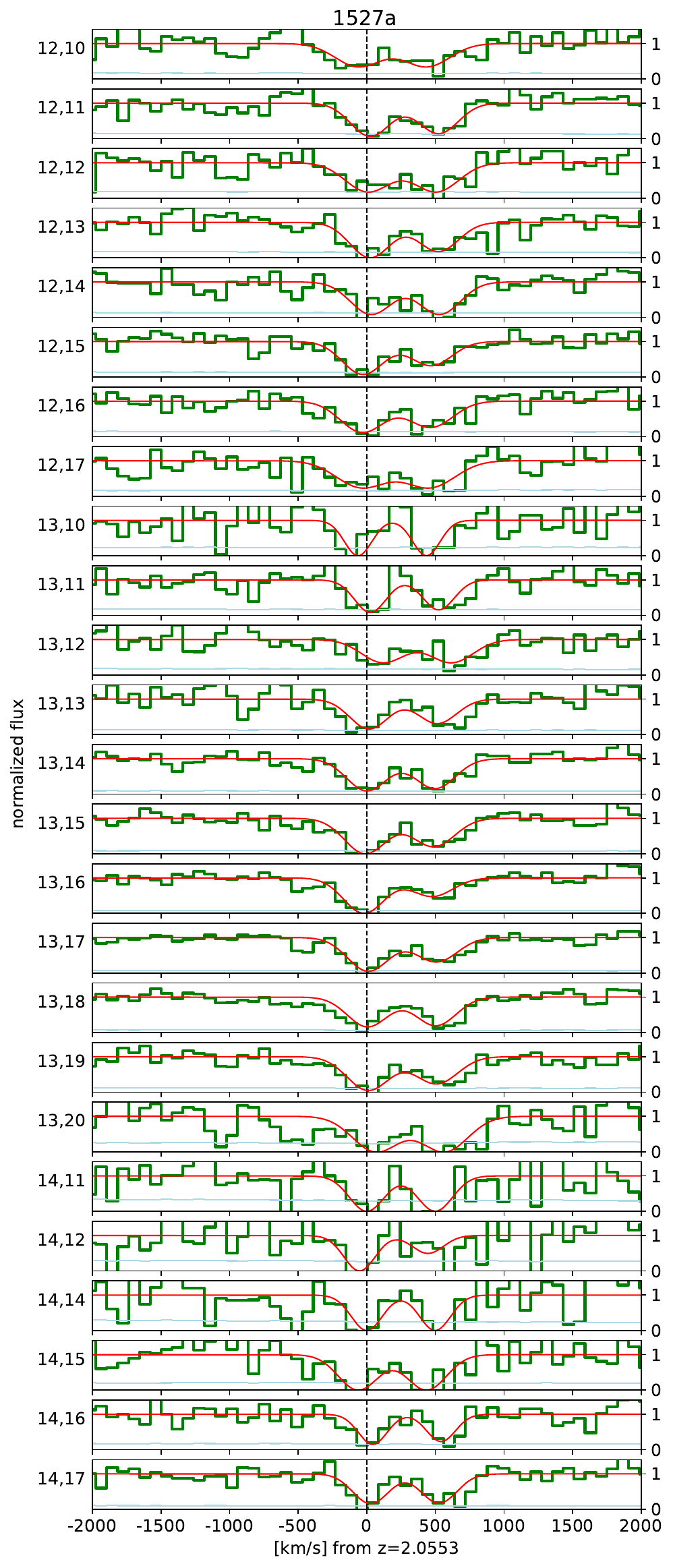}  
  \caption{\civ$\lambda\lambda 1548,1550$ absorption in system \one. The
    normalized flux is shown in green and the corresponding $1\sigma$
    error in light-blue. A double Gaussian fit is shown in red
    (\S~\ref{sect_fitting}). Each 
    panel corresponds to a binned arc spectrum having arbitrary spaxel 
    coordinates (cf. Fig.~\ref{fig_lensed}) indicated to the left
    side. {\it Figure continuing in the 
      next column.} 
  } \label{fig_data_1}
\end{figure}
\begin{figure}
  \ContinuedFloat
  \includegraphics[width=\columnwidth]{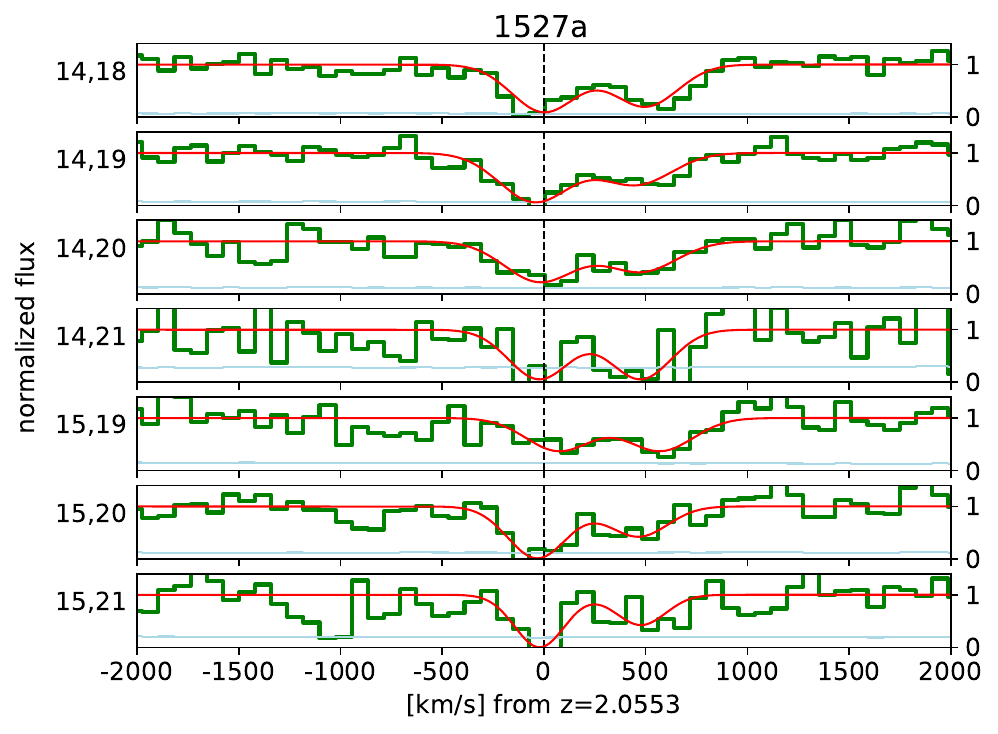}  
  \caption{
    Continued
  } 
\end{figure}

\begin{figure}
  \includegraphics[width=\columnwidth]{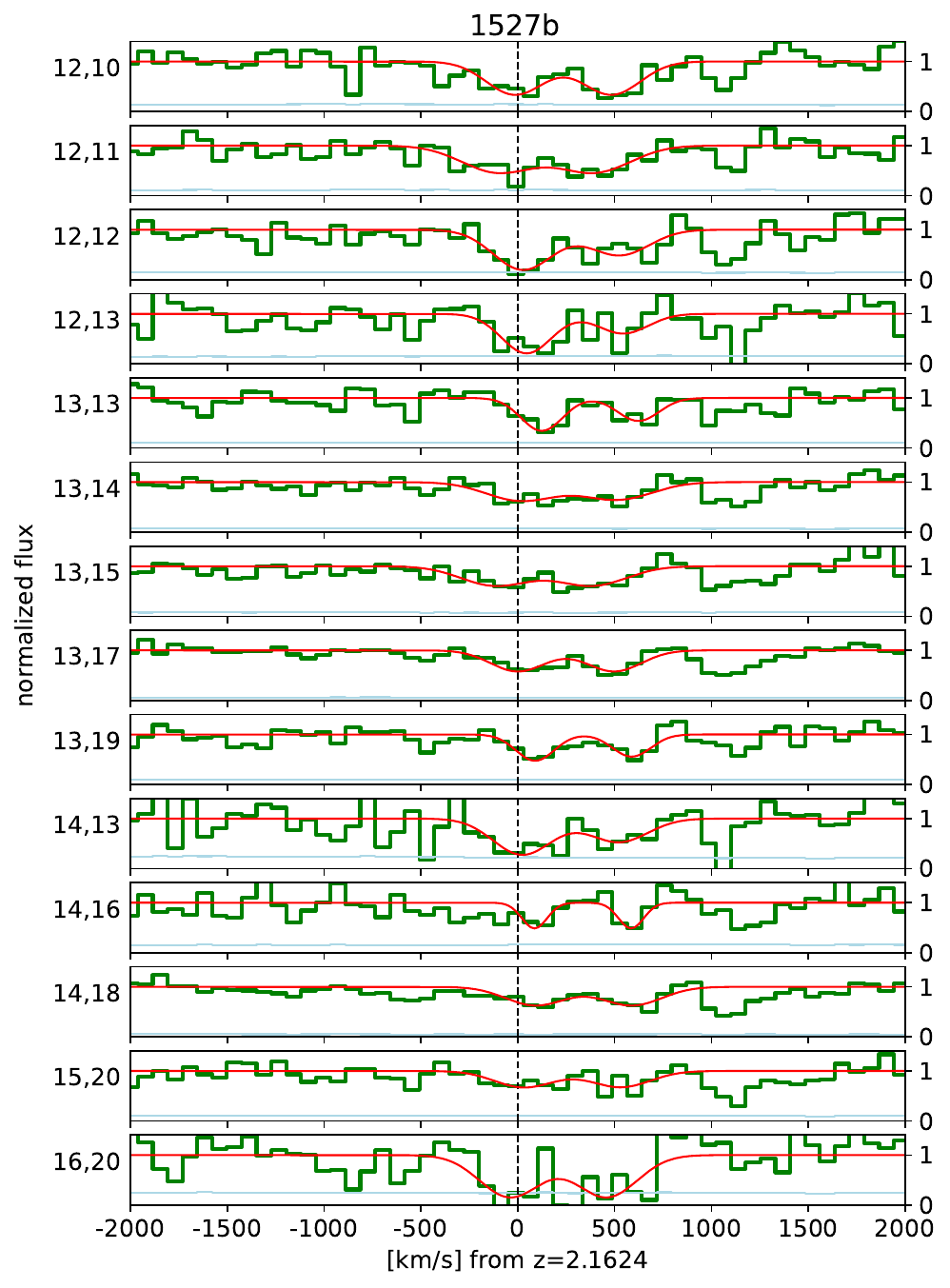}  
  \caption{Same as Fig.~\ref{fig_data_1} but for system \two. 
  } \label{fig_data_2}   
\end{figure}

\begin{figure}
  \includegraphics[width=\columnwidth]{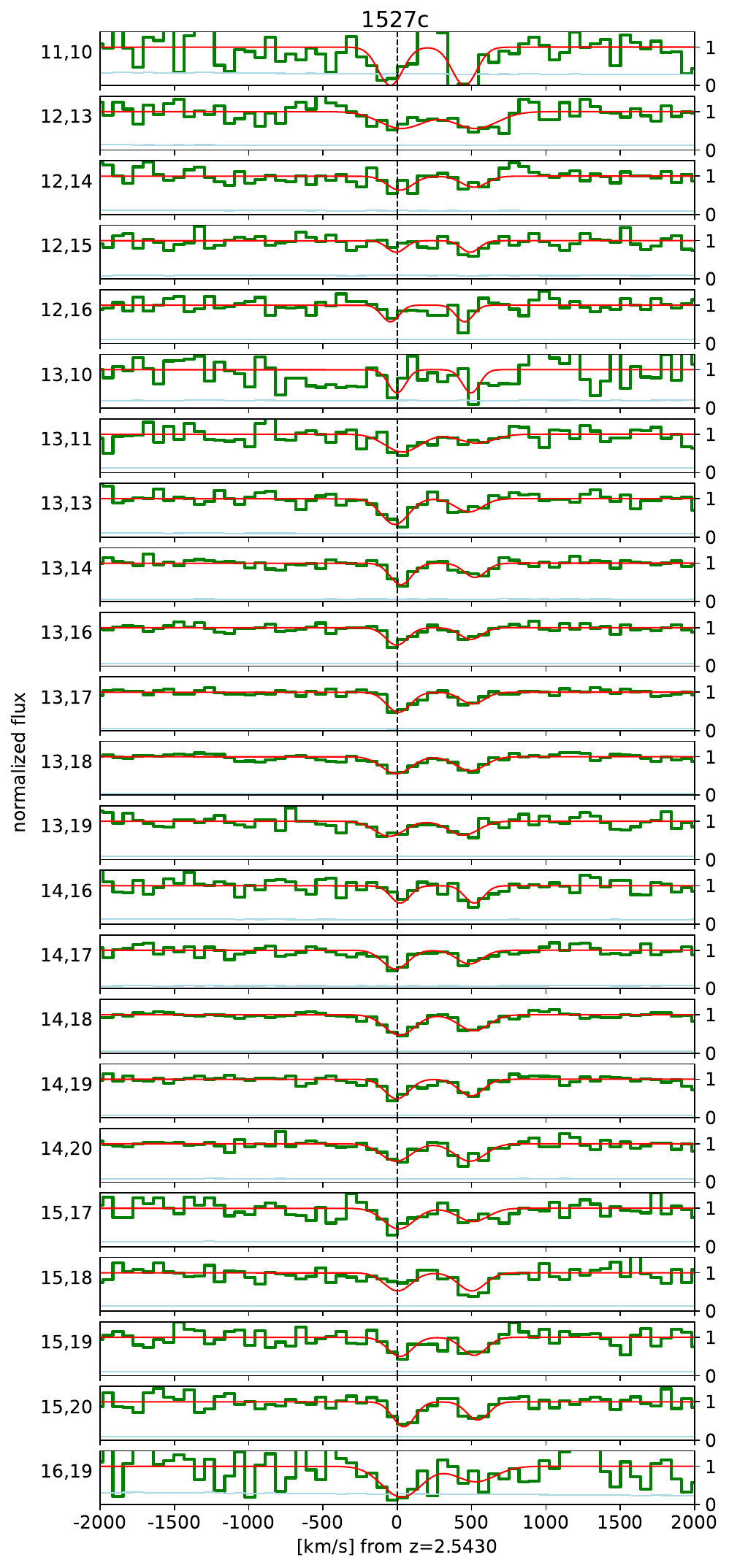}  
  \caption{Same as Fig.~\ref{fig_data_1} but for system \three.
  } \label{fig_data_3}
\end{figure}

\begin{figure}
  \includegraphics[width=\columnwidth]{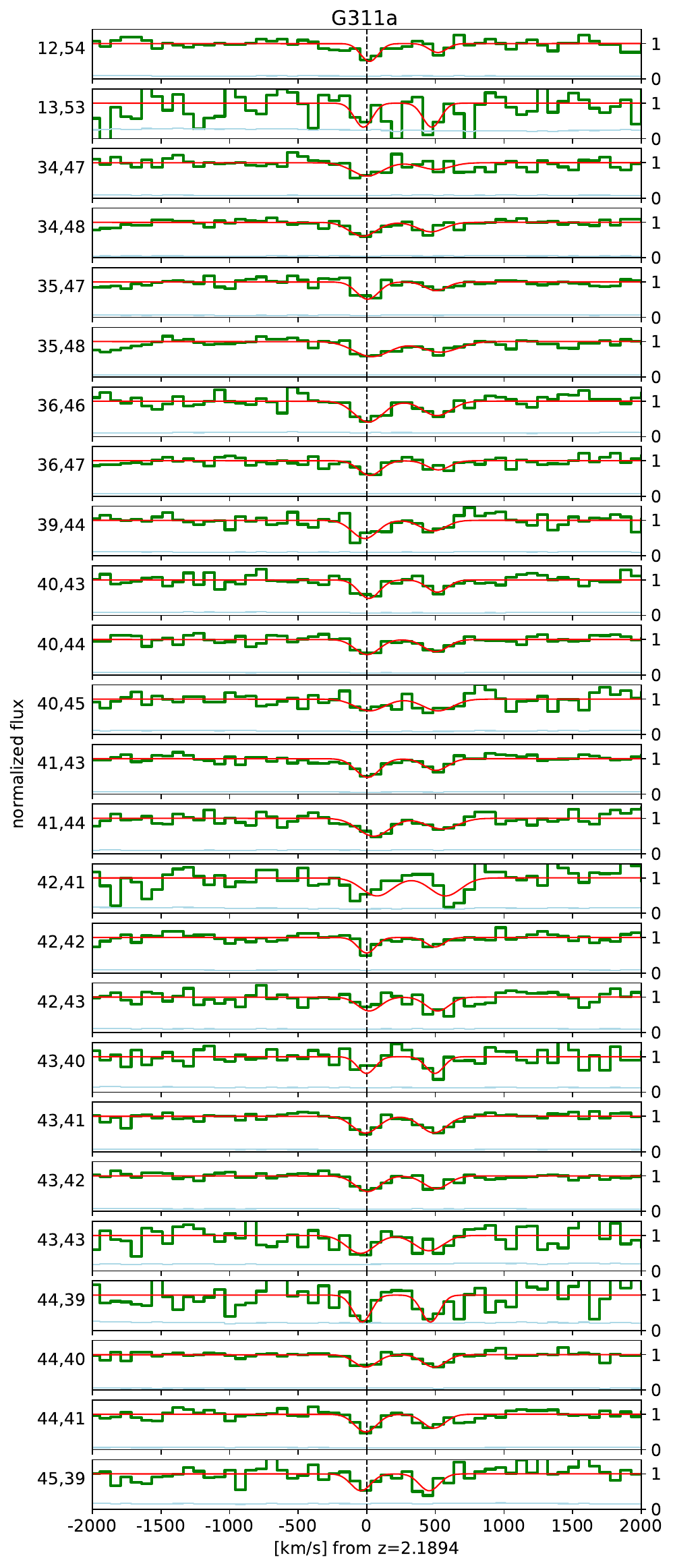}  
  \caption{Same as Fig.~\ref{fig_data_1} but for system \four. {\it
      Figure continuing in the next three columns.} 
  } \label{fig_data_4}
\end{figure}
\begin{figure}
  \ContinuedFloat
  \includegraphics[width=\columnwidth]{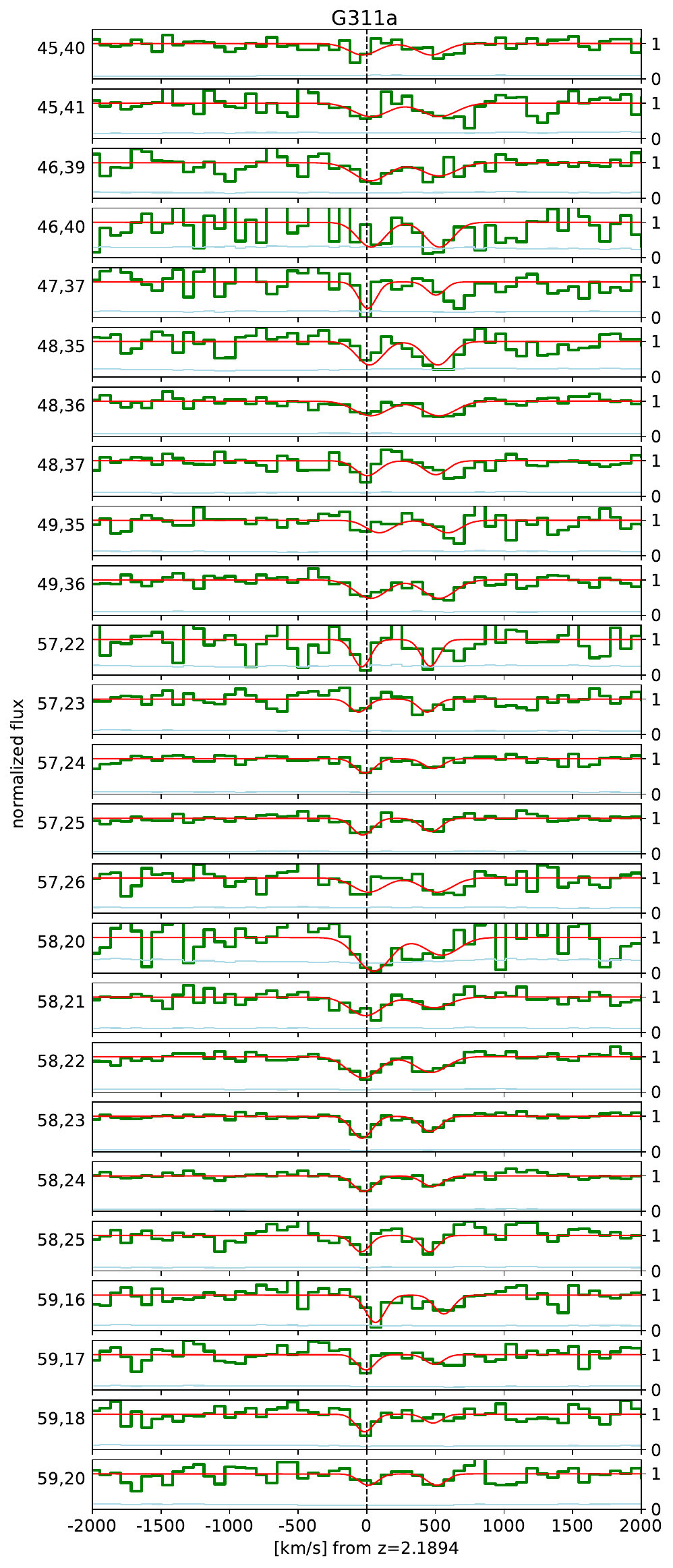}  
  \caption{
    Continued
  } 
\end{figure}
\begin{figure}
  \ContinuedFloat
  \includegraphics[width=\columnwidth]{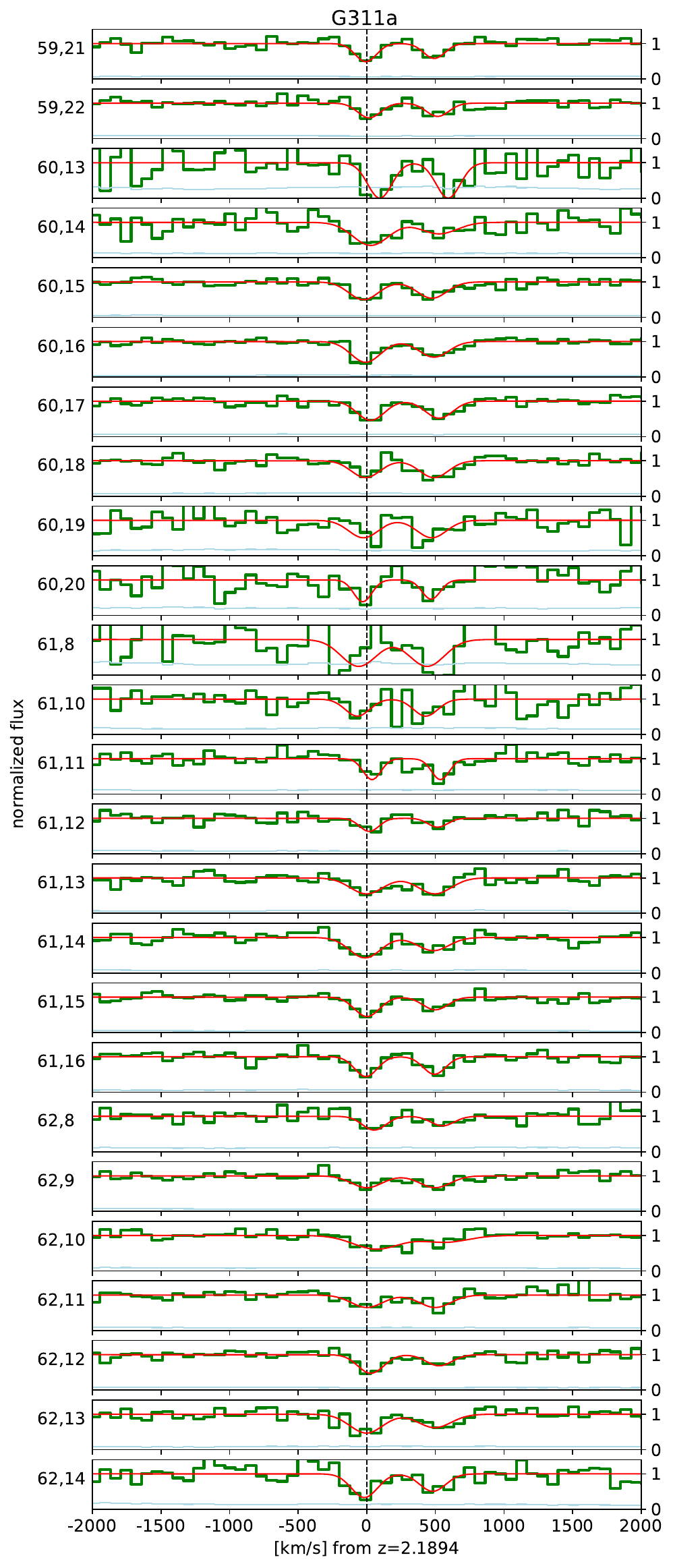}  
  \caption{
    Continued
  } 
\end{figure}
\begin{figure}
  \ContinuedFloat
  \includegraphics[width=\columnwidth]{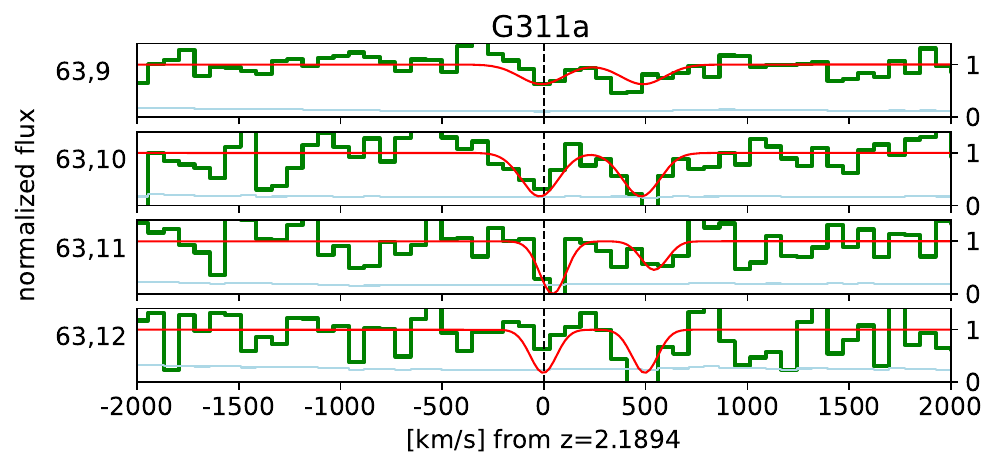}  
  \caption{
    Continued
  } 
\end{figure}

\begin{figure}
  \includegraphics[width=\columnwidth]{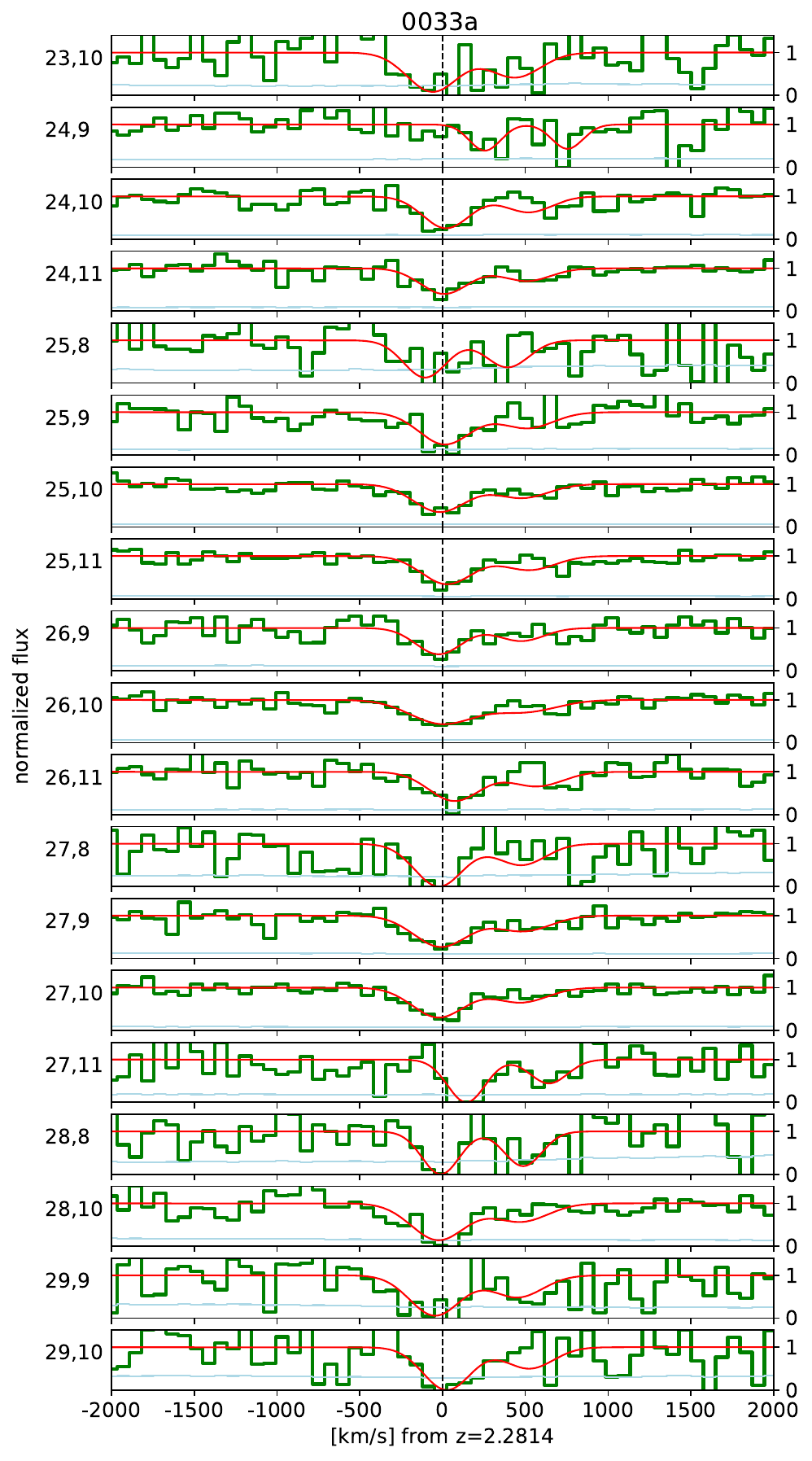}  
  \caption{Same as Fig.~\ref{fig_data_1} but for system \five.
  } \label{fig_data_5}
\end{figure}

\begin{figure}
  \includegraphics[width=\columnwidth]{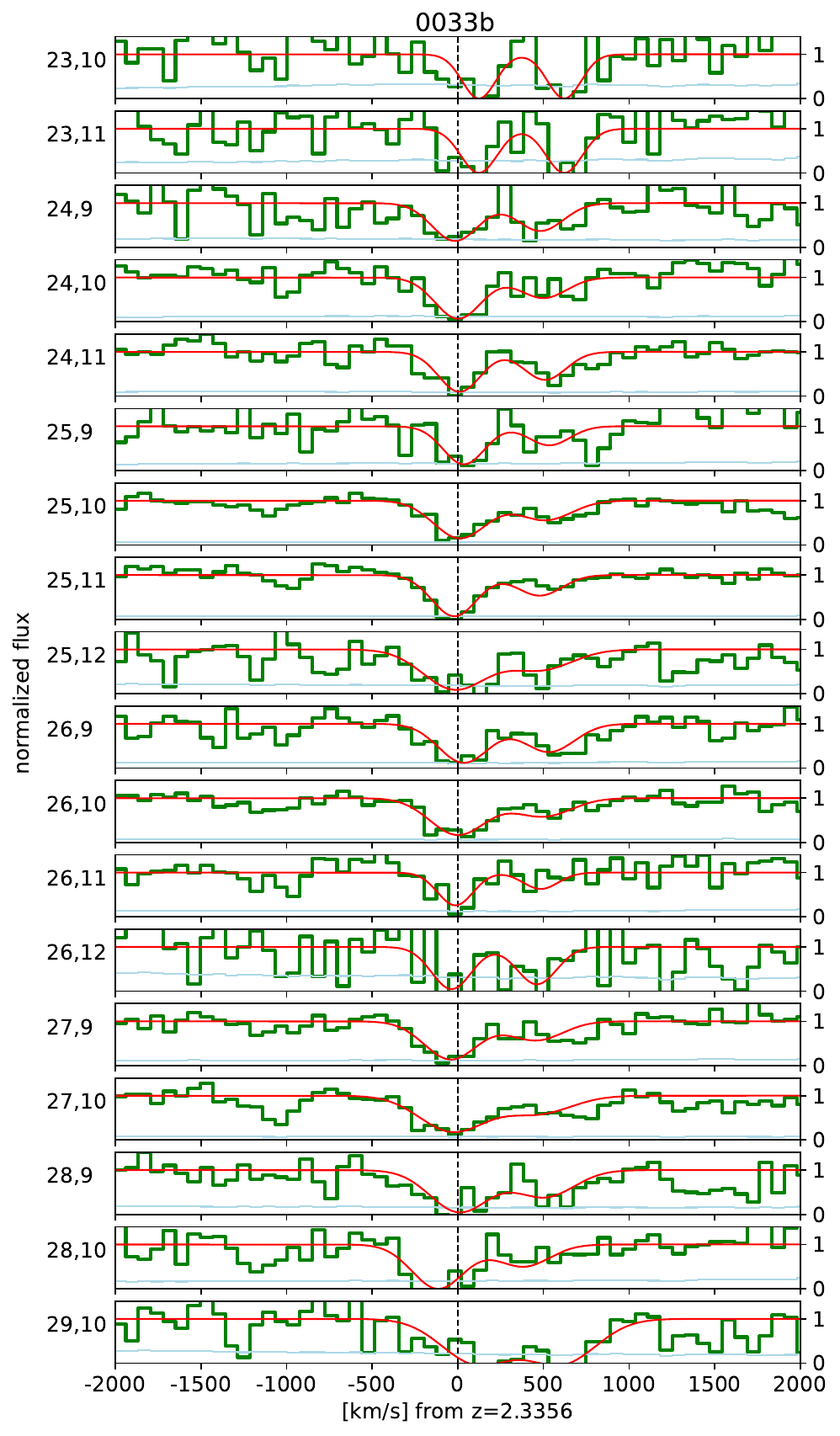}  
  \caption{Same as Fig.~\ref{fig_data_1} but for system \six.
    {\it  Figure continuing in the next column.} 
  } \label{fig_data_6}
\end{figure}

\begin{figure}
  \includegraphics[width=\columnwidth]{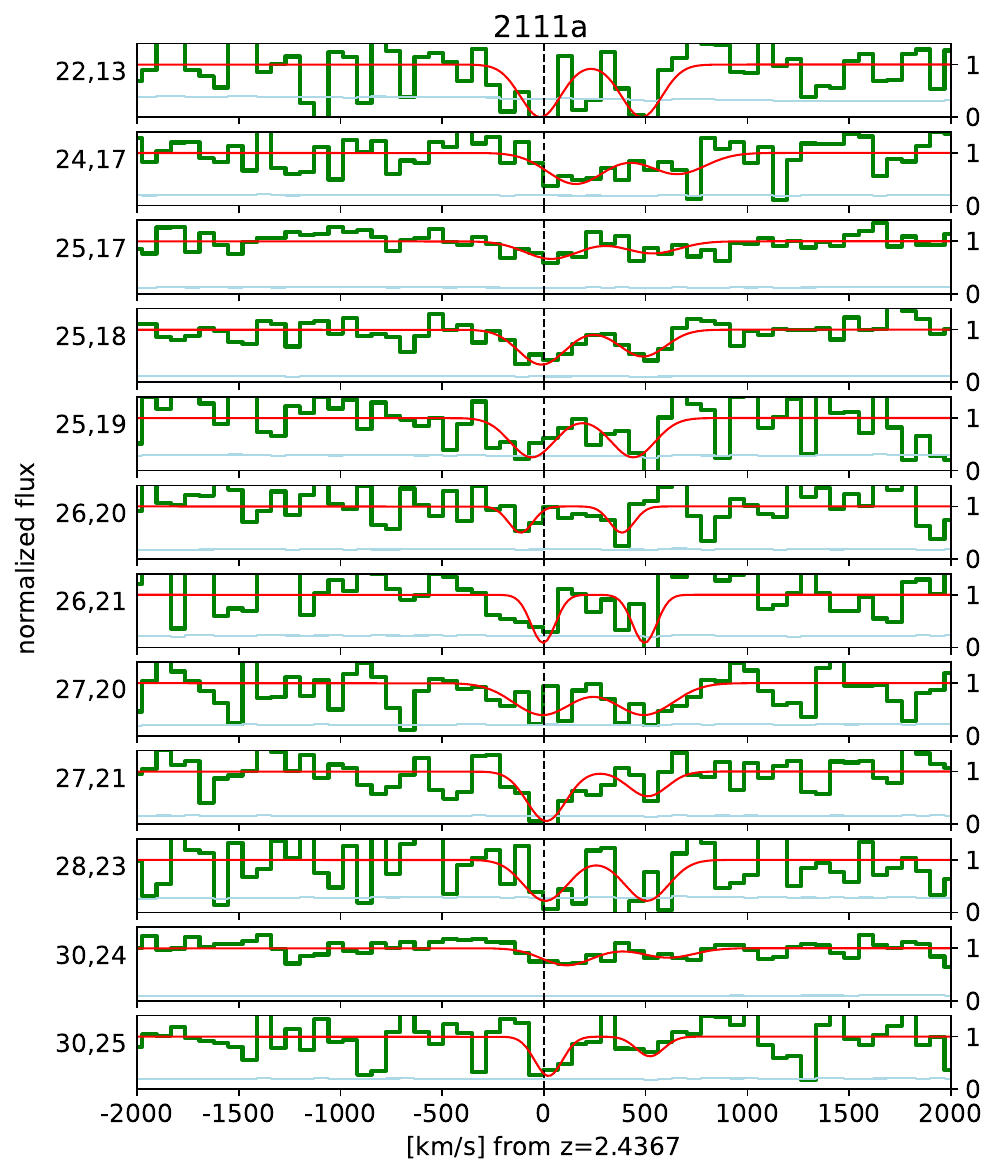}  
  \caption{Same as Fig.~\ref{fig_data_1} but for system \seven.
  } \label{fig_data_7}
\end{figure}

\begin{figure}
  \includegraphics[width=\columnwidth]{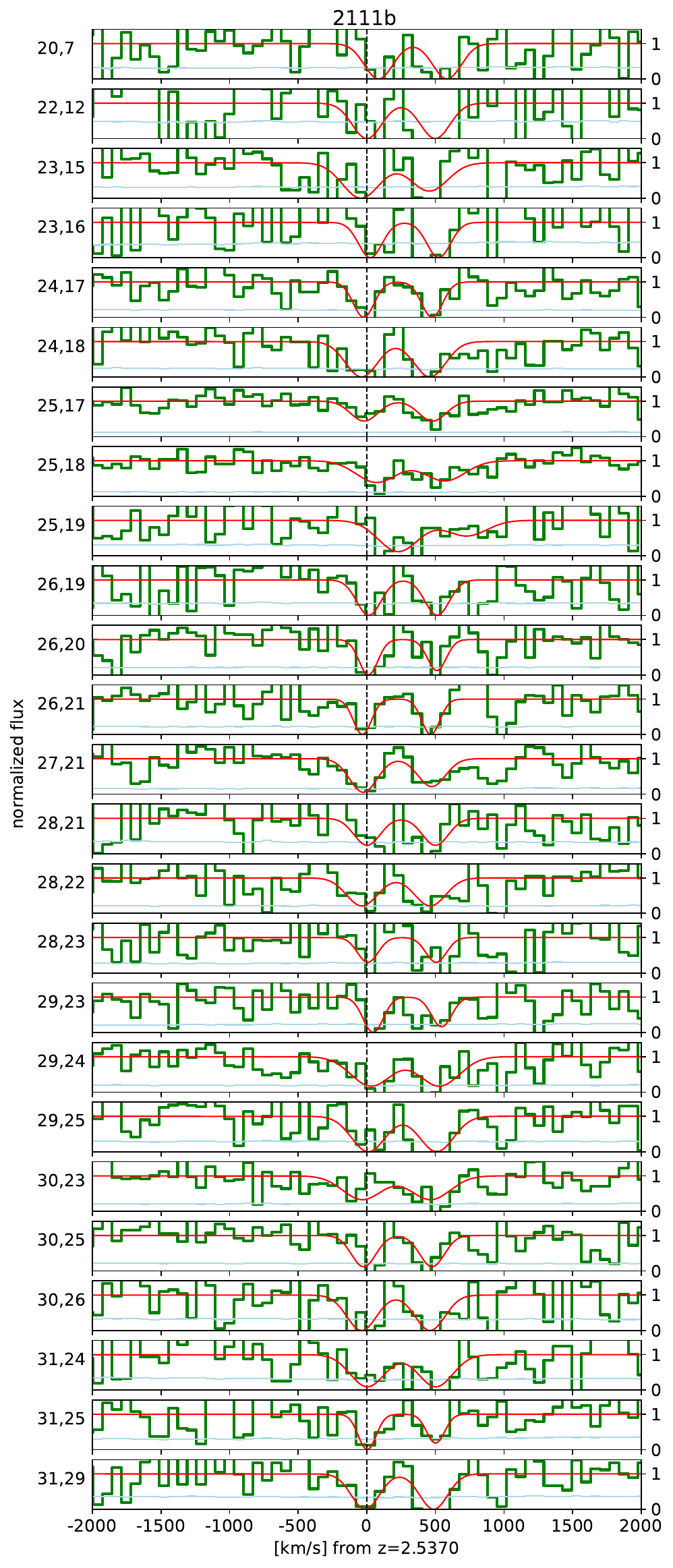}  
  \caption{Same as Fig.~\ref{fig_data_1} but for system \eight. {\it
      Figure continuing in the next column.}  
  } \label{fig_data_8}
\end{figure}

\clearpage
\tablefirsthead{\toprule
            Spaxel Coor. & $v$ & $\sigma_\text{obs}$ & $W_0$ \\
            $j, i$ & [km/s] & [km/s] & [$\mathrm{\AA}$]\\
            \midrule}
\tablehead{\repeatcaption\toprule
            Spaxel Coor. & $v$ & $\sigma_\text{obs}$ & $W_0$ \\
            $(j, i)$ & (km/s) & (km/s) & ($\mathrm{\AA}$)\\
            \midrule}
\topcaption{Gaussian fits results}\small
\begin{supertabular}[h]{cccc}
\midrule
\multicolumn{4}{l}{ \large \textbf{1527 a}\quad $z = 2.0552$} \\
\midrule
12,10 & $-61.7 \pm 28.2$ & $168.0 \pm 20.2$ & $1.42 \pm 0.28$  \\
12,11 & $24.7 \pm 13.9$ & $143.2 \pm 10.5$ & $1.73 \pm 0.2$  \\
12,12 & $7.8 \pm 24.0$ & $164.0 \pm 17.3$ & $1.74 \pm 0.3$  \\
12,13 & $23.5 \pm 15.6$ & $145.9 \pm 11.9$ & $1.89 \pm 0.24$  \\
12,14 & $31.9 \pm 13.8$ & $150.8 \pm 10.0$ & $1.79 \pm 0.2$  \\
12,15 & $-28.8 \pm 14.6$ & $148.9 \pm 11.1$ & $1.79 \pm 0.2$  \\
12,16 & $-32.4 \pm 18.0$ & $160.2 \pm 13.2$ & $1.83 \pm 0.23$  \\
12,17 & $-38.9 \pm 28.0$ & $183.9 \pm 20.2$ & $1.79 \pm 0.31$  \\
13,10 & $-63.6 \pm 15.2$ & $99.0 \pm 14.6$ & $1.28 \pm 0.28$  \\
13,11 & $25.1 \pm 13.4$ & $113.8 \pm 11.9$ & $1.35 \pm 0.21$  \\
13,12 & $118.7 \pm 26.1$ & $157.5 \pm 18.9$ & $1.33 \pm 0.26$  \\
13,13 & $15.7 \pm 13.9$ & $139.2 \pm 10.8$ & $1.52 \pm 0.18$  \\
13,14 & $-0.4 \pm 9.5$ & $147.3 \pm 7.0$ & $1.72 \pm 0.13$  \\
13,15 & $-10.7 \pm 8.7$ & $151.8 \pm 6.5$ & $1.96 \pm 0.13$  \\
13,16 & $-20.6 \pm 9.6$ & $147.5 \pm 7.6$ & $1.91 \pm 0.14$  \\
13,17 & $10.0 \pm 7.5$ & $150.5 \pm 5.7$ & $1.85 \pm 0.1$  \\
13,18 & $5.0 \pm 7.2$ & $146.7 \pm 5.4$ & $1.59 \pm 0.09$  \\
13,19 & $10.2 \pm 12.6$ & $153.5 \pm 9.4$ & $1.89 \pm 0.18$  \\
13,20 & $67.3 \pm 29.1$ & $170.5 \pm 20.8$ & $2.2 \pm 0.43$  \\
14,11 & $-0.3 \pm 24.3$ & $126.5 \pm 20.0$ & $1.63 \pm 0.41$  \\
14,12 & $-53.2 \pm 25.2$ & $113.3 \pm 22.7$ & $1.46 \pm 0.41$  \\
14,14 & $-0.5 \pm 17.6$ & $112.5 \pm 15.6$ & $1.45 \pm 0.32$  \\
14,15 & $-62.6 \pm 18.6$ & $144.6 \pm 14.0$ & $1.87 \pm 0.29$  \\
14,16 & $42.8 \pm 13.2$ & $104.9 \pm 12.4$ & $1.15 \pm 0.2$  \\
14,17 & $27.7 \pm 9.8$ & $129.2 \pm 7.8$ & $1.41 \pm 0.14$  \\
14,18 & $0.5 \pm 7.7$ & $158.9 \pm 5.7$ & $1.85 \pm 0.1$  \\
14,19 & $-45.4 \pm 10.1$ & $172.9 \pm 7.6$ & $2.08 \pm 0.13$  \\
14,20 & $-20.8 \pm 20.1$ & $172.9 \pm 14.8$ & $1.72 \pm 0.22$  \\
14,21 & $-23.5 \pm 28.6$ & $149.4 \pm 21.3$ & $1.82 \pm 0.42$  \\
15,19 & $73.2 \pm 24.3$ & $161.3 \pm 17.8$ & $1.31 \pm 0.24$  \\
15,20 & $-33.5 \pm 13.2$ & $143.5 \pm 10.5$ & $1.83 \pm 0.19$  \\
15,21 & $-21.7 \pm 18.5$ & $121.6 \pm 15.9$ & $1.57 \pm 0.29$  \\
\midrule
\multicolumn{4}{l}{ \large \textbf{1527 b}\quad $z = 2.1626$} \\
\midrule
12,10 & $-14.2 \pm 20.2$ & $148.0 \pm 15.1$ & $1.27 \pm 0.22$  \\
12,11 & $-103.4 \pm 28.3$ & $186.8 \pm 20.3$ & $1.29 \pm 0.23$  \\
12,12 & $26.6 \pm 21.2$ & $152.7 \pm 16.2$ & $1.58 \pm 0.25$  \\
12,13 & $45.6 \pm 19.7$ & $129.2 \pm 16.7$ & $1.32 \pm 0.25$  \\
13,13 & $124.0 \pm 13.1$ & $106.8 \pm 12.0$ & $0.91 \pm 0.15$  \\
13,14 & $15.5 \pm 24.1$ & $179.0 \pm 16.9$ & $0.87 \pm 0.13$  \\
13,15 & $-115.9 \pm 23.0$ & $177.4 \pm 16.7$ & $0.9 \pm 0.14$  \\
13,17 & $-0.4 \pm 12.2$ & $140.2 \pm 9.4$ & $0.77 \pm 0.09$  \\
13,19 & $89.3 \pm 12.8$ & $98.7 \pm 12.2$ & $0.67 \pm 0.12$  \\
14,13 & $23.4 \pm 34.5$ & $150.3 \pm 26.3$ & $1.41 \pm 0.37$  \\
14,16 & $86.4 \pm 16.8$ & $65.1 \pm 16.9$ & $0.43 \pm 0.17$  \\
14,18 & $90.8 \pm 13.7$ & $153.7 \pm 9.9$ & $0.74 \pm 0.08$  \\
15,20 & $32.7 \pm 31.5$ & $151.6 \pm 23.2$ & $0.65 \pm 0.17$  \\
16,20 & $-39.8 \pm 29.9$ & $157.4 \pm 21.9$ & $1.71 \pm 0.39$  \\
\midrule
\multicolumn{4}{l}{ \large \textbf{1527 c}\quad $z = 2.5430$} \\
\midrule
11,10 & $-46.4 \pm 16.8$ & $82.3 \pm 17.0$ & $1.06 \pm 0.33$  \\
12,13 & $28.0 \pm 24.7$ & $147.5 \pm 18.4$ & $0.84 \pm 0.17$  \\
12,14 & $17.6 \pm 17.4$ & $85.4 \pm 17.4$ & $0.4 \pm 0.12$  \\
12,15 & $-8.3 \pm 14.5$ & $60.9 \pm 14.9$ & $0.24 \pm 0.09$  \\
12,16 & $-46.4 \pm 11.7$ & $58.1 \pm 11.2$ & $0.33 \pm 0.1$  \\
13,10 & $-2.2 \pm 15.3$ & $58.1 \pm 15.8$ & $0.46 \pm 0.18$  \\
13,11 & $35.3 \pm 23.0$ & $117.9 \pm 20.2$ & $0.71 \pm 0.17$  \\
13,13 & $-16.3 \pm 10.4$ & $90.5 \pm 10.2$ & $0.79 \pm 0.12$  \\
13,14 & $23.8 \pm 7.4$ & $72.1 \pm 7.4$ & $0.52 \pm 0.07$  \\
13,16 & $-4.1 \pm 8.5$ & $73.3 \pm 8.6$ & $0.43 \pm 0.07$  \\
13,17 & $8.2 \pm 6.0$ & $79.3 \pm 6.0$ & $0.54 \pm 0.06$  \\
13,18 & $-5.8 \pm 6.6$ & $98.1 \pm 6.3$ & $0.56 \pm 0.05$  \\
13,19 & $-53.6 \pm 14.8$ & $100.9 \pm 14.0$ & $0.51 \pm 0.11$  \\
14,16 & $19.2 \pm 13.0$ & $64.7 \pm 12.8$ & $0.38 \pm 0.11$  \\
14,17 & $-12.9 \pm 9.5$ & $85.6 \pm 9.3$ & $0.54 \pm 0.08$  \\
14,18 & $16.0 \pm 5.9$ & $101.2 \pm 5.6$ & $0.69 \pm 0.05$  \\
14,19 & $-3.0 \pm 6.1$ & $81.2 \pm 6.0$ & $0.53 \pm 0.06$  \\
14,20 & $-9.0 \pm 12.3$ & $101.6 \pm 11.6$ & $0.6 \pm 0.11$  \\
15,17 & $10.7 \pm 17.6$ & $102.3 \pm 16.7$ & $0.71 \pm 0.16$  \\
15,18 & $2.8 \pm 16.7$ & $90.0 \pm 16.5$ & $0.55 \pm 0.15$  \\
15,19 & $18.2 \pm 12.4$ & $82.6 \pm 12.3$ & $0.53 \pm 0.12$  \\
15,20 & $42.7 \pm 8.9$ & $74.2 \pm 9.1$ & $0.63 \pm 0.11$  \\
16,19 & $29.7 \pm 33.7$ & $133.0 \pm 27.9$ & $1.38 \pm 0.42$  \\
\midrule
\multicolumn{4}{l}{ \large \textbf{G311 a}\quad $z = 2.1894$} \\
\midrule
12,54 & $19.9 \pm 11.2$ & $64.6 \pm 11.9$ & $0.42 \pm 0.1$  \\
13,53 & $-24.7 \pm 17.8$ & $64.6 \pm 18.3$ & $0.57 \pm 0.24$  \\
34,47 & $2.4 \pm 18.7$ & $109.8 \pm 17.6$ & $0.53 \pm 0.12$  \\
34,48 & $-29.2 \pm 10.0$ & $98.2 \pm 9.6$ & $0.5 \pm 0.07$  \\
35,47 & $6.3 \pm 8.7$ & $79.3 \pm 9.0$ & $0.5 \pm 0.08$  \\
35,48 & $38.3 \pm 11.9$ & $133.6 \pm 9.6$ & $0.74 \pm 0.08$  \\
36,46 & $12.8 \pm 14.6$ & $104.1 \pm 13.7$ & $0.8 \pm 0.15$  \\
36,47 & $23.2 \pm 11.4$ & $81.4 \pm 11.6$ & $0.44 \pm 0.09$  \\
39,44 & $-14.4 \pm 14.8$ & $92.2 \pm 15.0$ & $0.62 \pm 0.14$  \\
40,43 & $15.8 \pm 9.2$ & $75.7 \pm 9.3$ & $0.51 \pm 0.09$  \\
40,44 & $2.8 \pm 9.4$ & $88.1 \pm 9.6$ & $0.49 \pm 0.07$  \\
40,45 & $21.5 \pm 18.5$ & $108.3 \pm 16.8$ & $0.46 \pm 0.11$  \\
41,43 & $5.6 \pm 7.9$ & $86.0 \pm 8.0$ & $0.59 \pm 0.07$  \\
41,44 & $51.0 \pm 18.0$ & $113.9 \pm 16.3$ & $0.74 \pm 0.15$  \\
42,41 & $75.0 \pm 17.9$ & $110.3 \pm 16.0$ & $0.73 \pm 0.17$  \\
42,42 & $-0.8 \pm 12.1$ & $64.6 \pm 11.6$ & $0.37 \pm 0.09$  \\
42,43 & $18.0 \pm 15.1$ & $83.4 \pm 15.5$ & $0.42 \pm 0.11$  \\
43,40 & $-2.8 \pm 13.9$ & $64.6 \pm 14.0$ & $0.39 \pm 0.13$  \\
43,41 & $-9.9 \pm 7.4$ & $93.3 \pm 7.2$ & $0.57 \pm 0.07$  \\
43,42 & $7.0 \pm 10.2$ & $90.4 \pm 10.1$ & $0.52 \pm 0.08$  \\
43,43 & $-45.6 \pm 26.7$ & $101.5 \pm 25.3$ & $0.66 \pm 0.24$  \\
44,39 & $-34.4 \pm 14.3$ & $65.1 \pm 14.7$ & $0.64 \pm 0.21$  \\
44,40 & $-8.1 \pm 9.3$ & $82.2 \pm 9.3$ & $0.37 \pm 0.06$  \\
44,41 & $-8.2 \pm 6.9$ & $80.9 \pm 7.0$ & $0.52 \pm 0.06$  \\
45,39 & $-43.5 \pm 18.8$ & $72.6 \pm 19.1$ & $0.45 \pm 0.18$  \\
45,40 & $-28.6 \pm 20.9$ & $103.6 \pm 19.6$ & $0.43 \pm 0.12$  \\
45,41 & $30.1 \pm 33.1$ & $126.1 \pm 27.1$ & $0.62 \pm 0.21$  \\
46,39 & $27.8 \pm 27.7$ & $124.5 \pm 23.3$ & $0.83 \pm 0.23$  \\
46,40 & $31.0 \pm 27.2$ & $98.7 \pm 26.1$ & $0.89 \pm 0.36$  \\
47,37 & $5.2 \pm 15.2$ & $64.6 \pm 14.5$ & $0.62 \pm 0.18$  \\
48,35 & $19.9 \pm 20.7$ & $93.8 \pm 20.4$ & $0.8 \pm 0.26$  \\
48,36 & $29.0 \pm 16.0$ & $121.7 \pm 13.7$ & $0.65 \pm 0.12$  \\
48,37 & $4.0 \pm 16.0$ & $85.3 \pm 15.8$ & $0.47 \pm 0.13$  \\
49,35 & $91.7 \pm 21.6$ & $94.3 \pm 21.4$ & $0.43 \pm 0.15$  \\
49,36 & $29.7 \pm 14.3$ & $114.7 \pm 12.9$ & $0.77 \pm 0.14$  \\
57,22 & $-35.8 \pm 17.4$ & $64.6 \pm 17.8$ & $0.64 \pm 0.26$  \\
57,23 & $-58.9 \pm 14.6$ & $64.6 \pm 14.7$ & $0.31 \pm 0.11$  \\
57,24 & $-12.8 \pm 9.6$ & $70.8 \pm 9.6$ & $0.37 \pm 0.07$  \\
57,25 & $-29.9 \pm 7.7$ & $72.9 \pm 8.0$ & $0.45 \pm 0.07$  \\
57,26 & $14.4 \pm 29.3$ & $112.9 \pm 25.6$ & $0.59 \pm 0.21$  \\
58,20 & $50.0 \pm 30.0$ & $123.2 \pm 26.1$ & $1.51 \pm 0.45$  \\
58,21 & $-9.0 \pm 20.7$ & $117.0 \pm 18.9$ & $0.79 \pm 0.18$  \\
58,22 & $-27.2 \pm 10.4$ & $113.4 \pm 9.6$ & $0.87 \pm 0.1$  \\
58,23 & $-35.7 \pm 4.6$ & $73.4 \pm 4.9$ & $0.59 \pm 0.05$  \\
58,24 & $-19.7 \pm 8.6$ & $74.9 \pm 8.8$ & $0.43 \pm 0.07$  \\
58,25 & $-41.4 \pm 11.7$ & $64.6 \pm 12.1$ & $0.39 \pm 0.1$  \\
59,16 & $63.1 \pm 11.2$ & $69.5 \pm 11.1$ & $0.7 \pm 0.15$  \\
59,17 & $-5.5 \pm 14.4$ & $64.6 \pm 14.0$ & $0.36 \pm 0.11$  \\
59,18 & $-15.8 \pm 13.8$ & $64.6 \pm 13.2$ & $0.41 \pm 0.11$  \\
59,20 & $11.2 \pm 19.0$ & $69.8 \pm 19.1$ & $0.29 \pm 0.12$  \\
59,21 & $-7.0 \pm 7.8$ & $77.2 \pm 8.0$ & $0.51 \pm 0.07$  \\
59,22 & $17.5 \pm 10.6$ & $83.2 \pm 10.9$ & $0.45 \pm 0.08$  \\
60,13 & $94.7 \pm 18.9$ & $88.1 \pm 18.6$ & $1.14 \pm 0.36$  \\
60,14 & $29.0 \pm 19.6$ & $130.5 \pm 16.5$ & $1.1 \pm 0.2$  \\
60,15 & $-23.0 \pm 7.7$ & $107.7 \pm 7.2$ & $0.7 \pm 0.07$  \\
60,16 & $-8.5 \pm 6.0$ & $107.0 \pm 5.7$ & $0.81 \pm 0.06$  \\
60,17 & $28.5 \pm 7.8$ & $99.0 \pm 7.5$ & $0.69 \pm 0.08$  \\
60,18 & $-8.9 \pm 11.2$ & $104.4 \pm 10.3$ & $0.65 \pm 0.1$  \\
60,19 & $-24.5 \pm 20.0$ & $107.0 \pm 18.6$ & $0.68 \pm 0.19$  \\
60,20 & $-29.5 \pm 17.7$ & $66.1 \pm 18.1$ & $0.53 \pm 0.21$  \\
61,8 & $-59.2 \pm 33.7$ & $127.6 \pm 27.4$ & $1.25 \pm 0.44$  \\
61,10 & $-68.8 \pm 23.8$ & $89.1 \pm 23.5$ & $0.55 \pm 0.21$  \\
61,11 & $38.0 \pm 10.4$ & $64.6 \pm 10.6$ & $0.49 \pm 0.12$  \\
61,12 & $18.0 \pm 13.3$ & $70.0 \pm 13.4$ & $0.34 \pm 0.09$  \\
61,13 & $-2.0 \pm 13.5$ & $119.4 \pm 11.4$ & $0.7 \pm 0.1$  \\
61,14 & $-13.3 \pm 10.8$ & $112.9 \pm 9.8$ & $0.84 \pm 0.1$  \\
61,15 & $1.3 \pm 7.5$ & $91.2 \pm 7.5$ & $0.65 \pm 0.07$  \\
61,16 & $-5.8 \pm 7.0$ & $82.2 \pm 7.0$ & $0.6 \pm 0.07$  \\
62,8 & $53.8 \pm 20.3$ & $85.3 \pm 20.4$ & $0.43 \pm 0.15$  \\
62,9 & $0.1 \pm 12.5$ & $111.9 \pm 11.1$ & $0.48 \pm 0.07$  \\
62,10 & $69.0 \pm 28.3$ & $167.2 \pm 22.2$ & $0.83 \pm 0.16$  \\
62,11 & $0.6 \pm 16.4$ & $114.5 \pm 14.5$ & $0.52 \pm 0.1$  \\
62,12 & $29.1 \pm 9.7$ & $93.8 \pm 9.6$ & $0.64 \pm 0.09$  \\
62,13 & $-3.0 \pm 15.4$ & $120.4 \pm 13.4$ & $0.83 \pm 0.13$  \\
62,14 & $-17.7 \pm 15.2$ & $94.6 \pm 14.7$ & $0.82 \pm 0.18$  \\
63,9 & $-11.2 \pm 21.7$ & $108.8 \pm 19.6$ & $0.53 \pm 0.14$  \\
63,10 & $-20.5 \pm 12.4$ & $91.5 \pm 12.1$ & $0.98 \pm 0.19$  \\
63,11 & $43.4 \pm 11.3$ & $64.6 \pm 11.9$ & $0.83 \pm 0.2$  \\
63,12 & $-2.8 \pm 15.7$ & $64.6 \pm 15.5$ & $0.68 \pm 0.24$  \\
\midrule
\multicolumn{4}{l}{ \large \textbf{0033 a}\quad $z = 2.2813$} \\
\midrule
23,10 & $-60.0 \pm 28.6$ & $153.7 \pm 22.1$ & $1.82 \pm 0.39$  \\
24,9 & $255.7 \pm 21.1$ & $93.9 \pm 20.6$ & $0.74 \pm 0.24$  \\
24,10 & $21.5 \pm 14.0$ & $140.4 \pm 11.6$ & $1.35 \pm 0.16$  \\
24,11 & $9.2 \pm 16.7$ & $143.6 \pm 13.6$ & $1.11 \pm 0.15$  \\
25,8 & $-105.6 \pm 34.3$ & $129.1 \pm 27.9$ & $1.45 \pm 0.47$  \\
25,9 & $13.9 \pm 20.8$ & $155.2 \pm 16.6$ & $1.51 \pm 0.23$  \\
25,10 & $-17.3 \pm 11.2$ & $158.5 \pm 9.0$ & $1.33 \pm 0.11$  \\
25,11 & $26.6 \pm 11.2$ & $153.4 \pm 9.0$ & $1.3 \pm 0.11$  \\
26,9 & $-23.8 \pm 19.0$ & $137.6 \pm 15.6$ & $1.09 \pm 0.18$  \\
26,10 & $-1.8 \pm 17.7$ & $192.6 \pm 14.3$ & $1.39 \pm 0.14$  \\
26,11 & $71.9 \pm 20.1$ & $154.4 \pm 15.8$ & $1.36 \pm 0.19$  \\
27,8 & $-22.7 \pm 26.4$ & $145.4 \pm 22.1$ & $1.88 \pm 0.4$  \\
27,9 & $-10.9 \pm 18.6$ & $163.0 \pm 14.6$ & $1.54 \pm 0.19$  \\
27,10 & $-24.5 \pm 12.9$ & $157.7 \pm 10.3$ & $1.42 \pm 0.13$  \\
27,11 & $143.6 \pm 14.9$ & $113.0 \pm 13.8$ & $1.46 \pm 0.25$  \\
28,8 & $-11.4 \pm 23.9$ & $113.3 \pm 21.3$ & $1.46 \pm 0.4$  \\
28,10 & $-25.9 \pm 21.6$ & $163.2 \pm 17.1$ & $1.81 \pm 0.27$  \\
29,9 & $-45.4 \pm 31.6$ & $152.7 \pm 24.9$ & $1.86 \pm 0.44$  \\
29,10 & $26.1 \pm 32.3$ & $143.9 \pm 26.1$ & $1.86 \pm 0.48$  \\
\midrule
\multicolumn{4}{l}{ \large \textbf{0033 b}\quad $z = 2.3356$} \\
\midrule
23,10 & $123.2 \pm 18.9$ & $97.6 \pm 18.0$ & $1.26 \pm 0.36$  \\
23,11 & $123.0 \pm 18.5$ & $105.7 \pm 17.3$ & $1.36 \pm 0.34$  \\
24,9 & $-16.4 \pm 19.4$ & $134.1 \pm 15.3$ & $1.48 \pm 0.26$  \\
24,10 & $1.9 \pm 12.9$ & $134.6 \pm 10.6$ & $1.62 \pm 0.18$  \\
24,11 & $10.2 \pm 8.7$ & $122.0 \pm 7.4$ & $1.43 \pm 0.13$  \\
25,9 & $35.4 \pm 17.1$ & $120.8 \pm 15.1$ & $1.34 \pm 0.23$  \\
25,10 & $12.7 \pm 8.3$ & $153.4 \pm 6.4$ & $1.69 \pm 0.1$  \\
25,11 & $-20.7 \pm 7.0$ & $130.2 \pm 5.9$ & $1.57 \pm 0.1$  \\
25,12 & $-17.1 \pm 30.5$ & $185.3 \pm 23.5$ & $2.16 \pm 0.39$  \\
26,9 & $34.6 \pm 13.7$ & $146.7 \pm 10.1$ & $1.67 \pm 0.18$  \\
26,10 & $-0.2 \pm 9.9$ & $163.0 \pm 7.9$ & $1.75 \pm 0.12$  \\
26,11 & $-14.3 \pm 14.5$ & $103.0 \pm 13.6$ & $0.99 \pm 0.18$  \\
26,12 & $-36.7 \pm 24.8$ & $115.9 \pm 21.2$ & $1.43 \pm 0.41$  \\
27,9 & $-41.2 \pm 16.3$ & $154.9 \pm 12.8$ & $1.72 \pm 0.2$  \\
27,10 & $-30.6 \pm 14.8$ & $192.7 \pm 11.6$ & $2.02 \pm 0.17$  \\
28,9 & $9.3 \pm 21.4$ & $170.9 \pm 15.6$ & $2.09 \pm 0.28$  \\
28,10 & $-115.2 \pm 20.3$ & $152.2 \pm 16.3$ & $1.97 \pm 0.3$  \\
29,10 & $107.3 \pm 30.6$ & $202.1 \pm 21.7$ & $2.61 \pm 0.45$  \\
\midrule
\multicolumn{4}{l}{ \large \textbf{2111 a}\quad $z = 2.4368$} \\
\midrule
22,13 & $-17.8 \pm 21.6$ & $98.8 \pm 20.6$ & $1.28 \pm 0.41$  \\
24,17 & $156.0 \pm 31.3$ & $138.1 \pm 24.5$ & $1.05 \pm 0.28$  \\
25,17 & $35.7 \pm 32.0$ & $130.7 \pm 26.1$ & $0.57 \pm 0.17$  \\
25,18 & $-11.9 \pm 12.9$ & $112.9 \pm 11.5$ & $0.97 \pm 0.15$  \\
25,19 & $-59.6 \pm 24.4$ & $107.2 \pm 22.5$ & $1.04 \pm 0.34$  \\
26,20 & $-114.9 \pm 17.8$ & $59.9 \pm 16.5$ & $0.39 \pm 0.16$  \\
26,21 & $-4.0 \pm 11.2$ & $59.9 \pm 12.2$ & $0.71 \pm 0.21$  \\
27,20 & $-7.1 \pm 29.8$ & $142.7 \pm 22.8$ & $1.12 \pm 0.29$  \\
27,21 & $11.6 \pm 12.7$ & $95.4 \pm 12.5$ & $1.16 \pm 0.2$  \\
28,23 & $7.3 \pm 24.1$ & $107.7 \pm 22.3$ & $1.08 \pm 0.35$  \\
30,24 & $113.5 \pm 26.3$ & $123.0 \pm 22.8$ & $0.52 \pm 0.14$  \\
30,25 & $22.7 \pm 16.7$ & $67.9 \pm 16.8$ & $0.66 \pm 0.22$  \\
\midrule
\multicolumn{4}{l}{ \large \textbf{2111 b}\quad $z = 2.5370$} \\
\midrule
20,7 & $85.9 \pm 20.2$ & $103.0 \pm 18.9$ & $1.33 \pm 0.37$  \\
22,12 & $0.3 \pm 31.1$ & $106.7 \pm 28.9$ & $1.38 \pm 0.57$  \\
23,15 & $-42.8 \pm 27.3$ & $134.0 \pm 21.9$ & $1.73 \pm 0.43$  \\
23,16 & $25.3 \pm 23.3$ & $85.5 \pm 23.1$ & $1.1 \pm 0.44$  \\
24,17 & $-22.5 \pm 10.8$ & $76.2 \pm 10.9$ & $0.98 \pm 0.21$  \\
24,18 & $-38.7 \pm 15.8$ & $115.8 \pm 13.7$ & $1.5 \pm 0.28$  \\
25,17 & $-19.5 \pm 12.8$ & $99.2 \pm 12.3$ & $0.72 \pm 0.13$  \\
25,18 & $70.0 \pm 17.4$ & $147.7 \pm 12.8$ & $1.17 \pm 0.16$  \\
25,19 & $229.3 \pm 34.6$ & $146.1 \pm 28.2$ & $1.66 \pm 0.46$  \\
26,19 & $14.5 \pm 19.7$ & $88.3 \pm 19.3$ & $1.14 \pm 0.37$  \\
26,20 & $12.0 \pm 11.3$ & $68.3 \pm 11.2$ & $0.88 \pm 0.21$  \\
26,21 & $-32.1 \pm 11.0$ & $65.7 \pm 11.2$ & $0.85 \pm 0.21$  \\
27,21 & $-25.4 \pm 11.2$ & $99.9 \pm 10.7$ & $1.22 \pm 0.19$  \\
28,21 & $1.8 \pm 26.0$ & $94.8 \pm 24.9$ & $0.93 \pm 0.37$  \\
28,22 & $-35.9 \pm 17.8$ & $112.1 \pm 15.8$ & $1.15 \pm 0.25$  \\
28,23 & $6.8 \pm 21.1$ & $74.8 \pm 21.2$ & $0.68 \pm 0.28$  \\
29,23 & $46.2 \pm 13.2$ & $73.6 \pm 13.3$ & $0.95 \pm 0.24$  \\
29,24 & $30.8 \pm 19.7$ & $145.6 \pm 15.1$ & $1.56 \pm 0.26$  \\
29,25 & $11.0 \pm 20.6$ & $122.3 \pm 17.5$ & $1.58 \pm 0.36$  \\
30,23 & $-33.1 \pm 26.9$ & $142.8 \pm 20.5$ & $1.24 \pm 0.29$  \\
30,25 & $-23.0 \pm 13.4$ & $88.5 \pm 13.3$ & $1.01 \pm 0.23$  \\
30,26 & $-37.8 \pm 20.8$ & $107.9 \pm 18.9$ & $1.39 \pm 0.38$  \\
31,24 & $3.9 \pm 24.5$ & $124.6 \pm 20.0$ & $1.46 \pm 0.38$  \\
31,25 & $3.1 \pm 16.6$ & $64.8 \pm 16.8$ & $0.84 \pm 0.31$  \\
31,29 & $-11.2 \pm 21.4$ & $101.1 \pm 20.3$ & $1.31 \pm 0.4$  \\
\bottomrule
\end{supertabular}

\label{table_bestfit}

\section{On system identification and completeness  }
\label{sect_effectiveness}

The \civ\   system identification
  runs over all
  masked spectra along the whole available redshift path and provides
  redshift candidates that feed the 
  subsequent line profile 
  fitting (described
  in~Sect.~\ref{sect_fitting}).
The algorithm  employs Pearson correlation.  To 
begin, whole-spectrum continua are estimated through iterative
Savitzky-Golay filtering~\citep{Savitzky1964}. The identification is
performed along every continuum-normalized spectral pixel having S/N$
> 4$, wherein the observed flux is correlated with a template. The
template is built as two inverted Gaussian profiles defined by the
characteristic separation of the \civ\ doublet and the instrumental
spectral resolution. High correlation (Pearson coefficient $r > 0.7$,
typically) indicates a possible true
doublet~\citep[][]{Noterdaeme2010,Ledoux2015}. 
  The candidates are visually examined and classified
as a \civ\ system if high correlation occurs in at least one binned
spectrum.   
Eight systems are found  
(Table~\ref{table_summary}); 
notably, their $W_0$ distributions reach such high values that these 
systems would have been easily discovered by eye. 
This suggests we are not missing systems of this kind. 
To roughly
estimate  the system completeness { at the low-$W_0$ end}, we
inject  
synthetic \civ\ doublets with properties $z_i$ and $W_0$ and run 
the search in the vicinity of the $i$-pixel of each binned
spectrum. We obtain on average $\approx 50$\% 
recovery wherein { all} spectra in a system have a doublet with
$W_0\sim 0.7$ \AA.

\begin{figure}[H]
  \centering
  \includegraphics[width=\columnwidth]{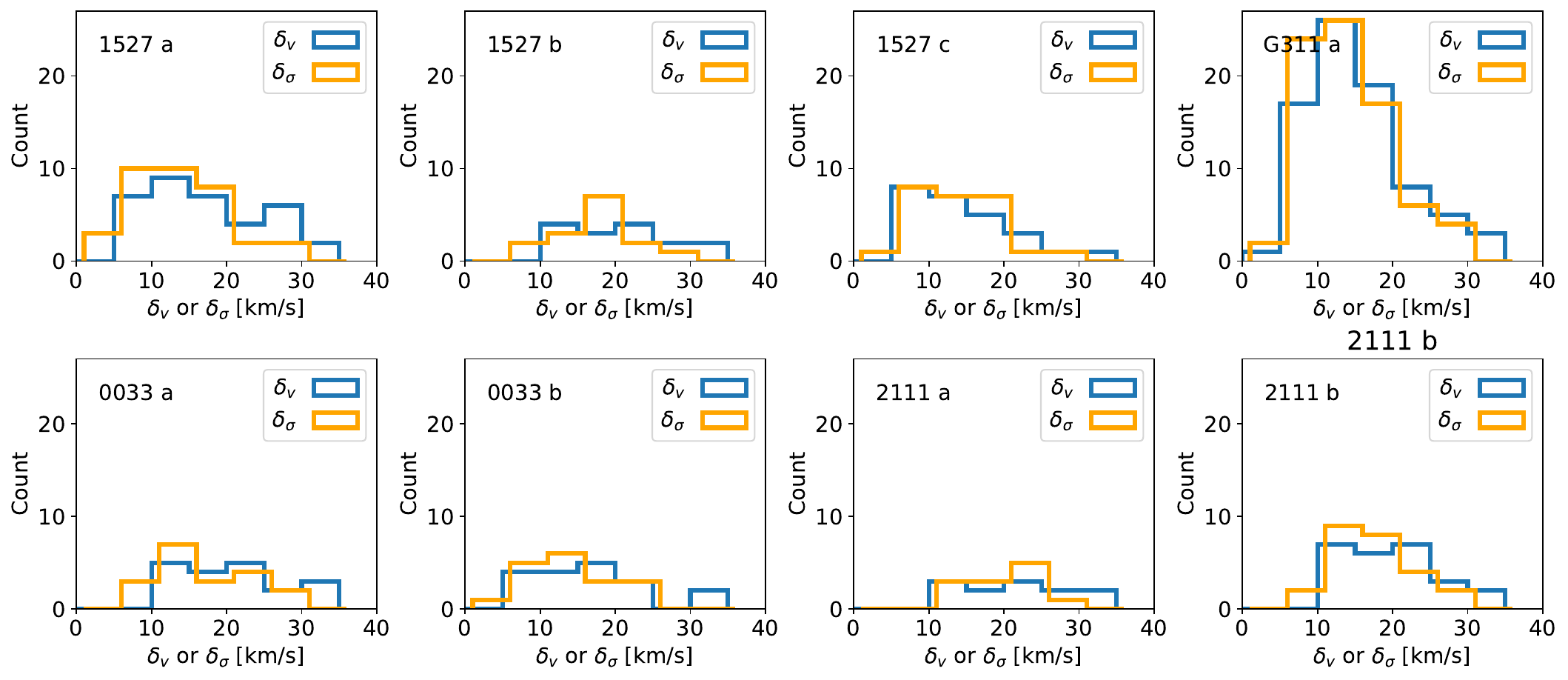}  
  \caption{Distribution of errors in centroid velocity ($\delta_v$) and absorption
    spread ($\delta_\sigma$).
  } \label{fig_errors}
\end{figure}

  \section{Possible spatial sampling biases}
  \label{sect_sampling}

To quantify spaxel-to-spaxel cross-talk effects, we performed two tests:

\begin{enumerate}[label={[\arabic*]}]

\item On the comparison between line-of-sight and transverse
    velocity dispersions (Sect.~\ref{sect_direct}): we re-extracted and
    re-fit spectra using a larger aperture of
    $0\farcs8\times0\farcs8$-binned spaxels. Consistently, this decreases
    the number 
    of spectra (by $\sim40\%$) but should also   
    counteract any possible cross-talk effects due to a
    FWHM$\approx 0\farcs8$ PSF. 
    The corresponding
    version of Fig.~\ref{fig_sigmas} shows broad consistency with the
    original one, that is, $\str< \spa$ holds. Therefore, this test indicates
    that the inequality is not an artifact induced
    by the chosen $0\farcs6\times0\farcs6$ aperture.

\item On the transverse auto-correlation of velocities
  (Sect.~\ref{sect_correlation}): a 40\% decrement in detections
  precludes the analysis of some of the systems due to their small 
  number of pairs. Instead, to quantify the possible effect
  of 
    $0\farcs6\times0\farcs6$ apertures on \tpcf\ we recompute it
  excluding all neighbor spaxels in the image plane; that is, pairs
  made of information from adjacent binned spaxels are not
  considered. This exercise removes $\sim10$\% of the sample pairs.
  Figure~\ref{fig_xi_1d_noadjacent} shows the results (for a better
  comparison we keep the mock and model curves as in
  Fig.~\ref{fig_xi_1d_all}).  In all systems the correlation signal,
  although noisier, is preserved, indicating it is not driven by pairs
  of adjacent spaxels. 
  Therefore, velocities measured using the
  $0\farcs6\times0\farcs6$ aperture can be considered independent, that is,
  not biased by spaxel 
  cross-talk effects. The exception is perhaps system \three, whose  
  correlation function is close to the signal expected from
  measurement errors only (indicated by the dashed lines in
  both figures). However, 
  this system also shows
  the strongest overlapping effect in the reconstructed absorber
  plane, so \tpcf\ could be biased toward $\Delta v=0$ \kms\ due to
  limitations of the lens model.

\end{enumerate}

\begin{figure}
  \centering
  \includegraphics[width=1\columnwidth]{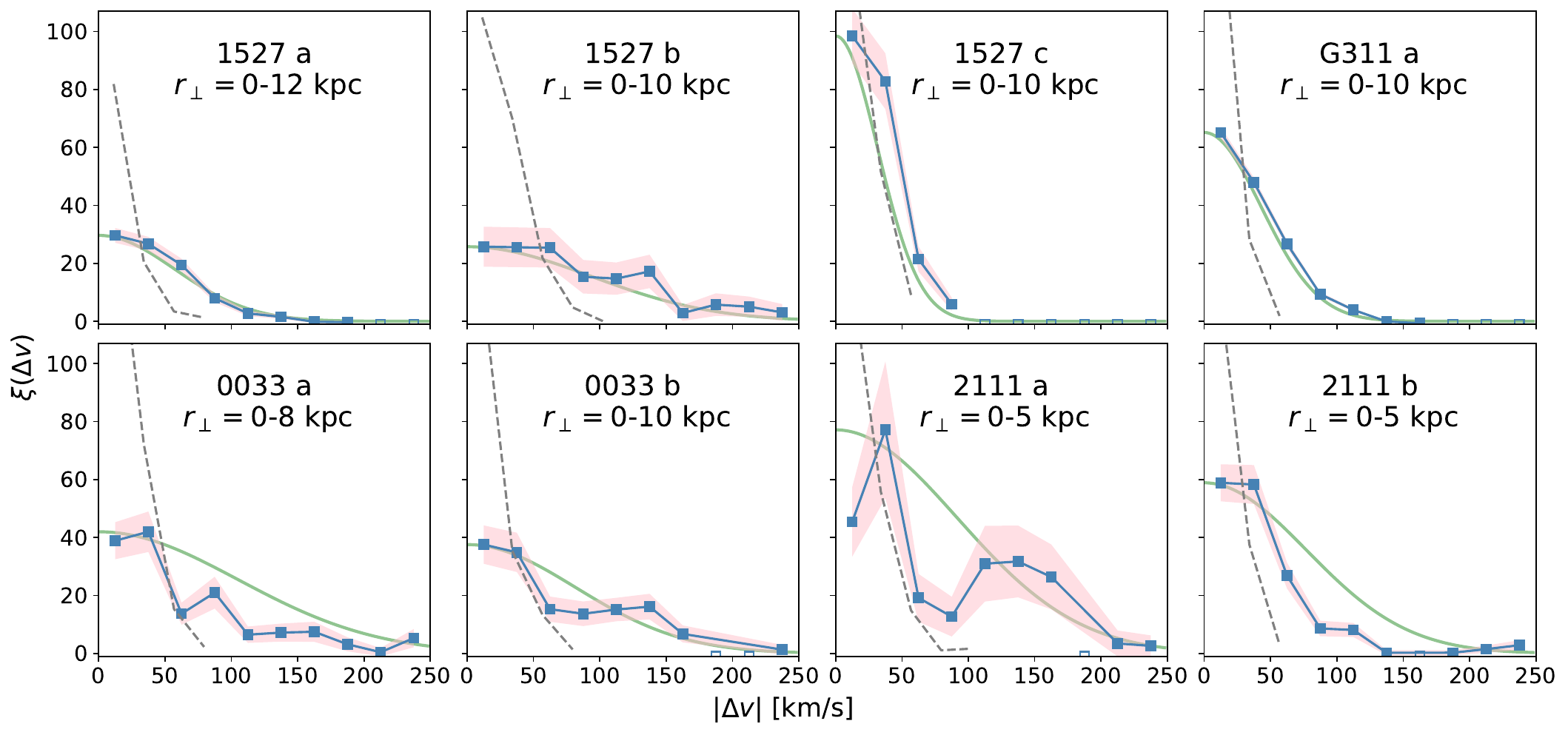}
  \caption{Same as Fig.~\ref{fig_xi_1d_all} but excluding pairs by
    neighbor spaxels in the image plane.}  
    \label{fig_xi_1d_noadjacent}. 
\end{figure}

  \section{Spurious transverse dispersion}
  \label{sect_spurious_dispersion}
  
  In~Sect.~\ref{sect_direct} we compare \str\ and $\langle \spa \rangle$. 
Naturally, we expect that low S/N spectra will lead
to an increased scatter in the velocity uncertainties measured from
Gaussian fitting, resulting in wider velocity distributions and thus
higher \str. In order to assess the magnitude of this systematic
effect, we perform a MC re-sampling test. Each detected \civ\ doublet
is ‘displaced’ to the same redshift, simulating a single-velocity
distribution. This is achieved by replacing the originally fitted
absorption lines with randomly selected flux from the adjacent
pixels. After the doublet is essentially erased a double Gaussian
profile is multiplied with the flux. These profiles have the same
properties as the absorption features found in the data, for the
corresponding spectra, except for the redshift which is set to the
defined velocity center of the system. \str\ 
obtained from fitting these doublets 
(Table~\ref{table_corrections})  should be purely due
to each system's S/N selection function.

\section{Sample dispersion approximation}
\label{sect_model}

Equation~\eqref{eq_long} is obtained by trial and error on data produced by
realizations of the model described in Sect.~\ref{sect_model_setup}.
  Figure~\ref{fig_sample_sigmas} shows realizations for
$\sigma_0=100$ \kms\ and six different number of clouds per 
spaxel, $N$. Each of these realizations   
deliver a sample dispersion ($\langle \spa \rangle$) mean and
median (blue and orange colors, respectively). The curves show solutions for the mean and the median (same respective colors). The 
exact solution for the mean is~\citep[][]{Kenney1951}:

\begin{equation}
\langle \spa \rangle_{mean} = \sqrt{\frac{2}{N}}~\frac{\Gamma{\left(
    \frac{N}{2}\right)}}{\Gamma{\left( \frac{N-1}{2}\right) }}~\sigma_0~, 
  \end{equation}

where $\Gamma(n)$ is the complete gamma function.
In this paper we use instead the approximation for the median given by
Eq.~\eqref{eq_long}. 

\begin{figure}[H]
  \centering
  \includegraphics[width=0.8\columnwidth]{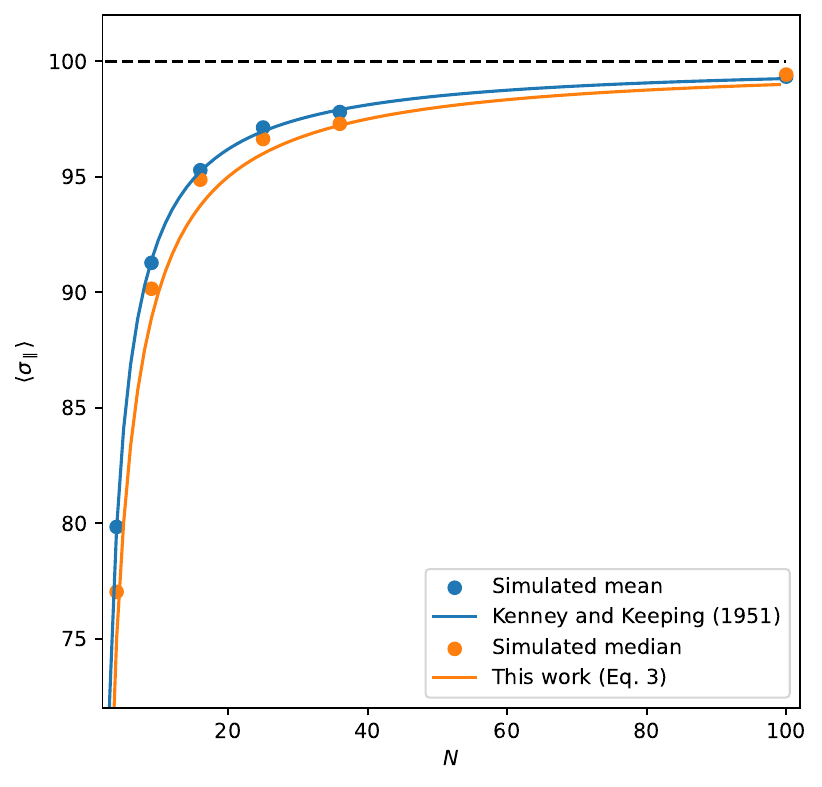}
  \caption{
    Mean and median sample dispersion versus number of
    scores per spaxel for $\sigma_0=100$ \kms. 
  } \label{fig_sample_sigmas}
  \end{figure}

\section{On the velocity auto-correlation function}

\subsection{Error budget}
\label{sect_error_xi}

{  
To account for the variance of \tpcf\ due to uncertainties in the
velocity measurements, $\sigma^2_{\xi}({\rm measurement})$, we ran 500
realizations of \tpcf\ by varying velocities according to ${\cal
  N}(v,\delta_{v}^2)$ and computed the variance of the median
statistics. 
On the other hand, the variance of \tpcf\ due to shot noise,
$\sigma^2_{\xi}({\rm Poisson})$,  was
computed according to 
\begin{equation}
\sigma^2_{\xi}({\rm Poisson}) = (DD/N_{DD}^2 + RR/N_{RR}^2) / (RR/N_{RR})^2~, 
\end{equation}  
where $DD$ ($RR$) is the number of data-data (random-random) pairs in a
($\Delta r_\perp,\Delta v$) bin and $N_{DD}$ ($N_{RR}$) is the total
number of respective pairs.   

The variance of \tpcf\ was taken to be the sum in quadrature of the
above variances. This is dominated by $ \sigma^2_{\xi}({\rm Poisson})$
due to both the small number statistics and the small velocity errors
compared to the velocity bins (Figs.~\ref{fig_xi_2d_all}
and~\ref{fig_xi_1d_all}).
}
\subsection{Random catalogs}
\label{sect_RRvelocities}

Sampling a uniform distribution ${\cal U}_{[a,b]}$ (of
variance $\sigma_{\cal U}^2=(b-a)^2/12$) results in a normal
distribution of variance $\sigma^2_{\cal N}=\sigma^2_{\cal U}/N$, where
$N$ is the number of averaged velocities. This implies 
that even a uniform velocity field will show velocity correlation if
recorded with a beam that does not resolve the clouds spatially.
Taking advantage of this property, we draw random ARCTOMO velocities
from ${\cal N}(0,\sigma_{RR}^2)$, where the random variance
$\sigma_{RR}^2 = (8\,000^2/12)/N$, that is, $\sigma_{RR}^2$ is the {\it
  binned} variance of ${\cal U}_{[-4000,4000]}$. For some systems we
extend this range until convergence is achieved.

\subsection{Mock catalogs}
\label{sect_mock}

\begin{figure*}

\subfloat[]{\includegraphics[width=0.66\textwidth]{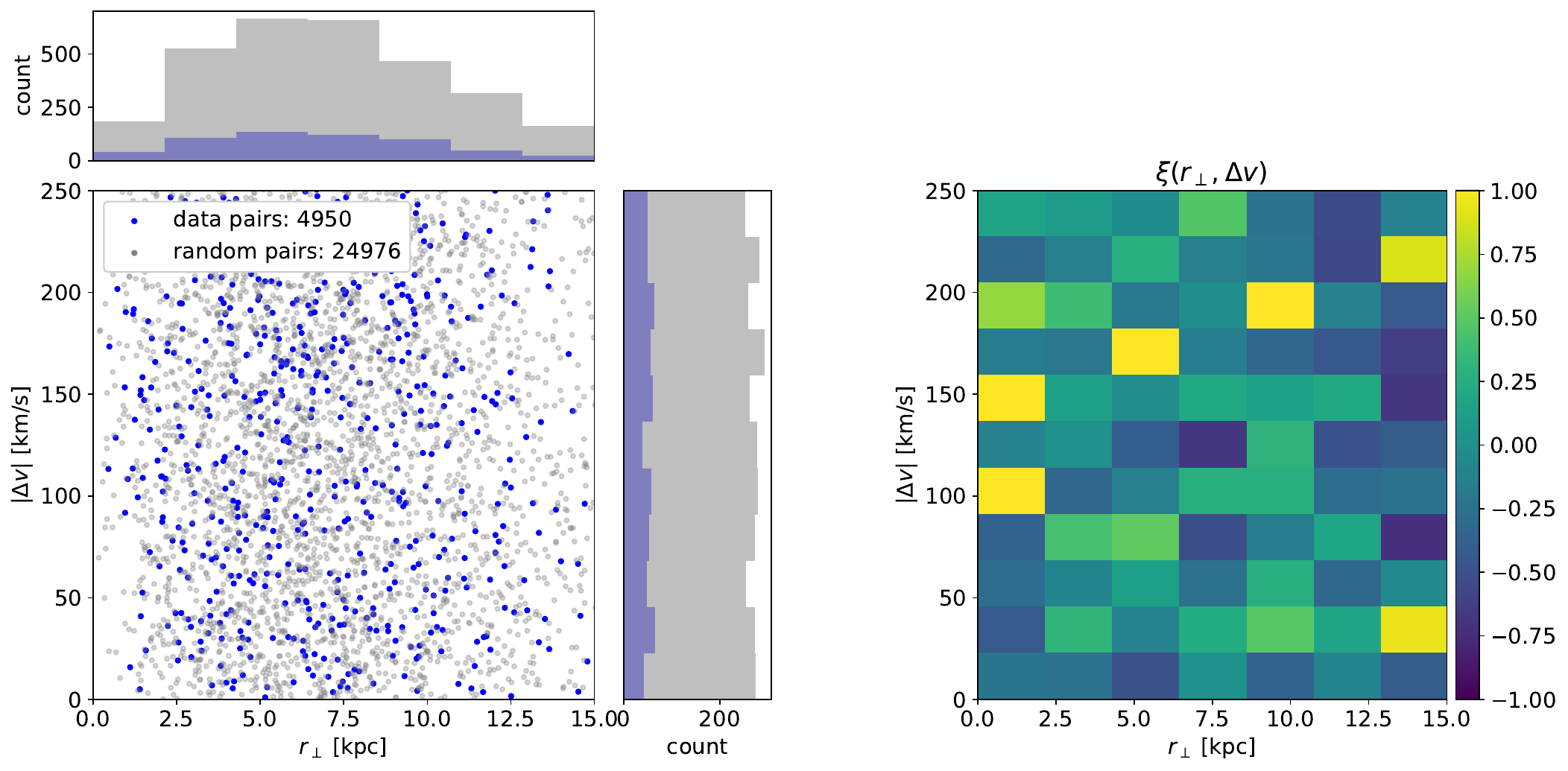} } 
\subfloat[]{\includegraphics[width=0.33\textwidth]{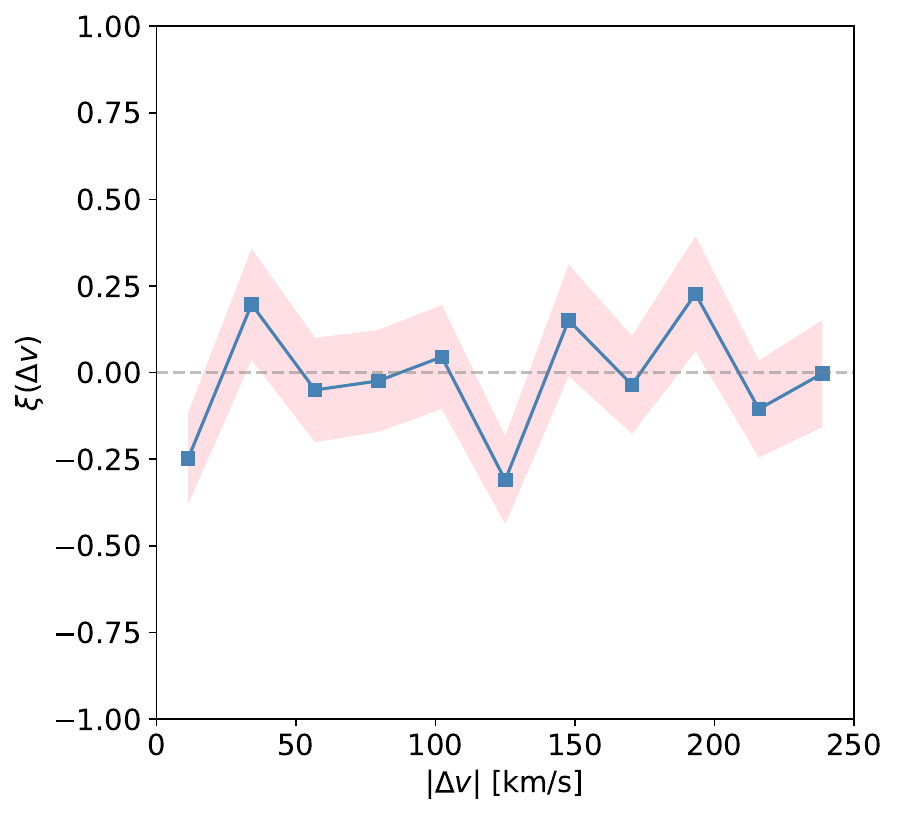}}
  \caption{\tpcf\  measured
    for a realistic 
    mock data set with no correlation. (a) In the scatter plot 
    the blue (gray) dots correspond to the mock data (random)
    pairs, and the side 
    panels show the respective projections in number counts. In the
    2d-histogram the color scale indicates \tpcf\ between -1 and 1. 
    (b) Velocity projection of \tpcf.
        The shaded pink regions indicate $1\sigma$ errors. 
  } \label{fig_mock}
\end{figure*}

Despite the safeguards to account for each system's complex
selection function, some limitations remain: (1) random
non-detections change the polygon edges~\citep{Hamilton2004} and thus
bias $\xi$ close to the edges in the spatial direction; (2)
individual data errors (both in velocity and spatially) can add
spurious signal in \tpcf.

To address these issues we build realistic mock catalogs by creating
re-sized replicas of the data samples, exactly like
in~Sect.~\ref{sect_random}, but 
having a random uniform $v$ in 
the range $(-4\,000,+4\,000)$ \kms, random RA-DEC normally
distributed around spaxel centers with standard deviation $2.4$
kpc (two de-lensed spaxels), and random $W_0$ drawn from the 
quasar $W_0$ distribution~\citep{Cooksey2013}. As in Sect.~\ref{sect_random},
detections and non-detections are defined by confronting $W_0$ with
the detection limit set by the S/N. Each mock entry passes the same
S/N criterion outlined above 
Fig.~\ref{fig_mock} shows \tpcf\ measured for a mock
data set, based on system \one. As expected, there is no significant
clustering power 
at any scale, which lends support to the
technique and 
also to the fidelity of the random catalogs.

  On the other hand, to re-create the spurious signal that a single
  velocity across spaxels would produce, mock catalogs are created as
  above but having velocities distributed as ${\cal N}(0,\delta_v^2)$,
  where $\delta_v$ are the individual measurement errors. The dashed
  curves in Fig.~\ref{fig_xi_1d_all} show the resulting \tpcf.  

\begin{figure}[H]
  \centering
\includegraphics[width=0.8\columnwidth]{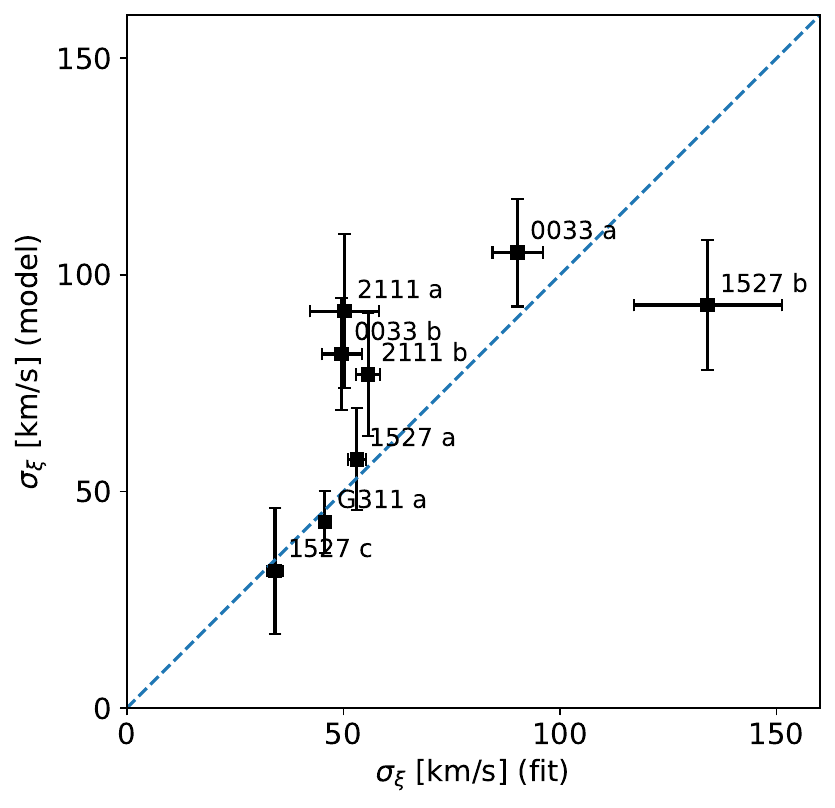} 
\caption{Per-system 1-$\sigma$ width of \xia\ displayed in
  Fig.~\ref{fig_xi_1d_all}. This is  
  computed via a single Gaussian fit of \xia\ ($x$-axis) or 
  predicted by the kinematic model described by Eq.~\eqref{eq_xi}, 
  using $\sigma_0$ 
  and   $N$ in Table~\ref{table_sigmas} ($y$-axis). Error bars 
  represent fit and propagated errors, respectively. 
  } \label{fig_xis}
\end{figure}

\end{appendix}
\end{document}